\documentclass{aastex61}
\usepackage{natbib}

\usepackage{amsmath}
\usepackage{multirow}
\usepackage{graphics}
\usepackage{hyperref}
\usepackage{txfonts}
\usepackage{enumitem}
\usepackage{tikz}
\usetikzlibrary{calc}
\usetikzlibrary{arrows, decorations.markings}

\begin{document}


\title{3D magneto-hydrodynamic models of non-thermal photon emission in the binary system $\gamma^2$~Velorum}

\author{K.~Reitberger, R.~Kissmann, A.~Reimer, O.~Reimer}
\affil{Institut f\"ur Astro- und Teilchenphysik and Institut f\"ur Theoretische Physik, Leopold-Franzens-Universit\"at Innsbruck, A-6020 Innsbruck, Austria} 
\email{klaus.reitberger@uibk.ac.at}



\begin{abstract}
Recent reports claiming association of the massive star binary system $\gamma^2$~Velorum (WR 11) with a high-energy $\gamma$-ray source observed by \textit{Fermi}-LAT contrast the so-far exclusive role of $\eta$~Carinae as the hitherto only detected $\gamma$-ray emitter in the source class of particle-accelerating colliding-wind binary (CWB) systems. We offer support to this claim of association by providing dedicated model predictions for the nonthermal photon emission spectrum of $\gamma^2$~Velorum. \\

We use three-dimensional magneto-hydrodynamic modeling (MHD) to investigate the structure and conditions of the wind-collision region (WCR) of $\gamma^2$~Velorum including the important effect of radiative braking in the stellar winds. A transport equation is then solved throughout the computational domain to study the propagation of relativistic electrons and protons. The resulting distributions of particles are subsequently used to compute nonthermal photon emission components.\\

In agreement with observation in X-ray spectroscopy, our simulations yield a large shock-cone opening angle. We find the nonthermal $\gamma$-ray emission of $\gamma^2$~Velorum to be of hadronic origin owing to the strong radiation fields in the binary system which inhibit the acceleration of electrons to energies sufficiently high for efficient inverse Compton radiation. We also discuss the strong dependence of a hadronic $\gamma$-ray component on the energy-dependent diffusion used in the simulations. Of two mass-loss rates for the WR star found in literature, only the higher one is able to accommodate the observed $\gamma$-ray spectrum with reasonable values for important simulation parameters such as the injection ratio of high-energy particles within the WCR. \\

\end{abstract}

\keywords{acceleration of particles -- binaries: general -- gamma rays: stars -- hydrodynamics -- stars: winds, outflows}


\section{INTRODUCTION} 
\label{intro}
The $\gamma^2$ Velorum (WR 11) binary system is of interest for many different reasons. At a distance of 336$^{+8}_{-7}$ pc \citep{North2007} it is the closest known binary system containing a WR~star \citep{Henley2005} and contains the brightest example of an WR type star with an O-star companion. The two stars combined reach 1.8 mag and are well visible with the naked eye at the Southern night sky. $\gamma^2$ Velorum, along with its early-type stellar companion $\gamma^1$ Velorum, is part of the Vela OB2 association, a group of $\sim$100 early-type stars spread over a diameter of $\sim 70$ pc at a mean distance of 410 $\pm$ 12 pc \citep{Jeffries2009}.\\

The spectral types of the two stellar components of WR~11 are most probable WC8 and O7.5 \citep{deMarco2000}. Both stars orbit each other with a period of 78.53 $\pm$ 0.01 days \citep{Schmutz1997} with an eccentricity of 0.334 $\pm$ 0.003 \citep{North2007}. From periastron to apastron the stellar separation of the two components varies from $\sim$172 to $\sim$344 R$_\odot$.
$\gamma^2$ Velorum is one of very few binary systems where direct determination of the stellar masses is possible and allows for critical tests of (WR and) massive star evolution. It constitutes an excellent laboratory for the study of radiatively-driven winds \citep{Millour2007}.\\

$\gamma^2$ Velorum is observable at many wavelengths. In the radio regime, it is the strongest known thermal source among the class of hot luminous stars \citep{Purton1982} and was first detected by \citet{Seaquist1976} using the Parkes 64m telescope. More recent analysis with ATCA gave strong indication for a highly attenuated nonthermal radio component originating at the WCR between the two stars \citep{Chapman1999}.  The attenuation of this nonthermal radio component is plausibly explained by the absorption of radio waves in the wind of the WR star \citep{Becker2013}. The observed indication for nonthermal radio waves via synchrotron emission suggests the presence of high-energy electrons accelerated in the WCR. \\

$\gamma^2$ Velorum was observed by ISO at 2.3 to 29.6 $\mathrm{\mu}$m \citep{vanderHucht1996} and by Keck in the K-band during an investigation of several WR systems among which $\gamma^2$ Velorum and WR~140 were the only two objects without any observable dust emission throughout the system and its WCR \citep{Monnier2002}. More recent efforts applying long-baseline interferometry in near-IR with the AMBER/VLTI instrument revealed that the observed infrared emission primarily results from the contribution of the individual stars of the binary system. The stellar separation is too small to allow to spatially resolve the two components. Discrepancies between models and data may be resolved by additional free-free emission originating in the WCR \citep{Millour2007}.\\ 

Observations at optical wavelengths are manifold. Most relevant for this study are perhaps the observations with the HEROS spectrograph at the ESO 50 cm telescope \citep{Schmutz1997,deMarco1999,deMarco2000} and its interpretation regarding system parameters, as well as more recent observations with the Sydney University Stellar Interferometer \citep{North2007} which set the so far best constraints on many important stellar and stellar wind parameters by a new determination of the orbital parameters of the system. Interestingly, the determination of the WR-star's mass-loss rate is ambiguous as different methods (polarimetric and radio-emission-based) yield different answers; the former yielding a mass-loss rate of $\sim$8.0$\times 10^{-6}$ M$_\odot$yr$^{-1}$, the latter of  $\sim$3.0$\times 10^{-5}$ M$_\odot$yr$^{-1}$. The difference might plausibly stem from the effect of wind clumping on the radio-emission-based mass-loss rate which is proportional to the square of the wind density. The polarimetrically determined mass-loss rate is proportional to the density and therefore insensitive to clumping \citep{North2007}.\\

From observations at ultra-violet frequencies some details concerning the structure of WCR in $\gamma^2$ Velorum can be derived. Using data from Copernicus and the International UltraViolet Explorer, \cite{St.-Louis1993} find clear evidence of eclipses of the O-star light caused by the WR-wind, as well as the presence of a wide cavity in the wind that is much closer to the O-star than to its companion. Thus, the first evidence for wind-wind collision in $\gamma^2$ Velorum came for UV data.\\

The role of wind clumping as well as constraints on the opening-angle of the wind cavity have been further explored by X-ray observations. $\gamma^2$ Velorum is a bright and well observable soft X-ray emitter and has been seen by ROSAT \citep{Willis1995}, ASCA \citep{Rauw2000}, Chandra \citep{Skinner2001}, and XMM-Newton \citep{Schild2004}. The latter study reveals a curious high and low state variation in the 1 to 8 keV regime that may originate from photoelectric absorption in the dense WR-wind whenever it eclipses the hot collisional plasma of the WCR. Whenever the WCR can be seen through a rarefied cavity that builds around and behind the O-star, the X-ray flux is clearly enhanced. \cite{Henley2005} use the measured X-ray variation in data provided by Chandra to estimate that the shock-cone opening half angle of the wind cavity must be rather large ($\sim$ 85$^\circ$) -- a finding which could not be reproduced in their hydrodynamical models which generally show a much narrower WCR. The same study concludes that the winds of both components will not reach terminal velocity before reaching collision. This puts additional constraints on hydrodynamical models as it requires more sophisticated wind implementations (e.g., radiative wind acceleration). 
The X-ray emission seen from $\gamma^2$ Velorum is attributed to thermal emission reaching up to $\sim8$~keV. Investigations at higher energies towards the $\gamma$-ray regime by INTEGRAL have yielded upper limits \citep{Tatischeff2004}. The same authors also derive an upper limit on a possible nonthermal component in the X-ray data seen by ASCA. \\

At high-energy $\gamma$ rays a recent study of 7 years of \textit{Fermi}-LAT data shows a weak $\gamma$-ray signal (6.1 $\sigma$) at 0.1 to 100 GeV at the position of the binary system \citep{Pshirkov2016}. Variability of the $\gamma$-ray flux consistent with $\gamma^2$ Velorum's orbital period could not yet be established due to low statistics. If this $\gamma$-ray source should indeed stem from the WCR in $\gamma^2$ Velorum, it would be the second object detected within the source class of particle accelerating CWB systems - next to $\eta$ Carinae \citep{Tavani2009,Takapaper,Reitberger2015}. \\

$\gamma^2$ Velorum's closeness to Earth and its well constrained stellar and stellar wind parameters make it a prime target for numerical modeling. The system's short orbit, low eccentricity and lack of complicated gas and dust structures (like the Homunculus nebula around the $\eta$ Carinae) simplify modeling efforts and increase the reliability of the results. \\

It is the aim of this study to present a MHD model of $\gamma^2$ Velorum which reproduces the above mentioned large shock-cone opening half angle derived from Chandra data and predicts a $\gamma$-ray signature to be compared with the observations by the \textit{Fermi}-LAT. Insights gained by our model may be used to further constrain model parameters (injection ratios, diffusion coefficients, etc.) needed for the modeling of further or even more complex systems. The ability to successfully model the observed $\gamma$-ray emission will also support the claim of association.\\

The numerical modeling framework we use to model $\gamma^2$ Velorum's wind evolution, its particle populations and nonthermal emission is based on the work presented in \cite{Reitberger2014}, \cite{Reitberger2014b}, and \cite{Kissmann2016}. As these studies have so far applied the code to generic binary systems, the present study describes the first application of the code to a real astrophysical system.
In Section \ref{newstuff} we detail and motivate all additions and alterations to the numerical modeling framework that have been implemented since the above mentioned publications, most notably the inclusion of ``strong coupling'' in the radiative wind-acceleration, an improved treatment for high-energy diffusion, and also for shock acceleration.
In Section \ref{results} we present our modeling results on the system of $\gamma^2$ Velorum and compare it to observations.
Finally, Section \ref{disc} provides a discussion of the results and an outlook on implications for future modeling of other CWB systems - most notable $\eta$ Carinae and WR~140.\\

\section{Updates of the numerical modeling framework} 
\label{newstuff}
\subsection{Radiative wind-acceleration}
Implementing stellar winds with a fixed terminal velocity does not reflect reality and is disfavored by observations showing that the winds in $\gamma^2$ Velorum hit each other long before reaching terminal velocity \citep{Henley2005}. A more sophisticated approximation of radiative wind-acceleration becomes necessary. Its details of implementation are a crucial point of our model as they significantly alter the outcome. In previous work we have relied on the standard Castor-Abbott-Klein (CAK) approximation \citep{CAK} in its modified form proposed by \cite{Pauldrach1986}. So far, we have used its variant of ``weak coupling''. The current version of the code can also use the case of ``strong coupling''. In the following we motivate the existence of these two approaches and argue about our choice of method for the $\gamma^2$ Velorum system.\\

A commonly used expression for the line acceleration of a single star is
\begin{align}
	\vec{g}_{rad,i}^l=g_{rad,i}^l \frac{\vec{r}_i}{r_i}
\end{align}
	with the radial acceleration term
\begin{align}
	g_{rad,i}^l = \frac{\sigma_e}{c} \frac{L_{\star,i}}{4\pi r^2}
	k t^{-\alpha} I_{FD}
\end{align}
the optical depth parameter
\begin{align}
	t = \left(\sigma_e \rho v_{th}\right)/\left|\frac{\partial u}{\partial r
} \right|
\end{align}
the finite-disc correction 
\begin{align}
	I_{FD} = \frac{1}{1-\cos \theta_{\star,i}^2} \frac{1}{\pi}
	\int_0^{2\pi}
	\int_0^{\theta_{\star,i}}
	\left(
	\left(\vec{n}_i\cdot \nabla\left(\vec{n}_i\cdot\vec{u}\right)\right)
		/ \left|\frac{\partial u}{\partial r}\right|
		\right)^{\alpha} \cos \theta_i d \Omega_{i}
\end{align}
and the angle
\begin{align}
\sin\theta_{\star,i}=\frac{R_{\star,i}}{r_i}
\end{align}


The radial acceleration component $g_{rad,i}^l$ is therefore a function of the distance to the star $r_i$, its bolometric Luminosity $L_{\star,i}$, the velocity of the wind $\vec{u}$, the stellar radius $R_{\star,i}$ and the standard CAK parameters $\alpha$ and $k$.
\begin{align}
g_{rad,i}^l=f(\vec{u},L_{\star,i},R_{\star,i},r_i,\alpha,k)
\label{lines}
\end{align}

If applied to a system of only one star, these equations can be readily applied. However, if we consider two stars, their interpretation is more difficult.
The effect of the companion star's radiation field on the primary star's wind, and vice versa, will lead to radiative inhibition (star B's radiation field reduces the wind acceleration in star A's atmosphere) and radiative braking (rapid deceleration of star A's wind - shortly before reaching the WCR - due to the increasing radiative force of star B). To address both effects correctly, one has to know how the radiation of star B acts on the wind of star A and vice versa. This, however, is not entirely clear as Eq. \ref{lines} leaves room for two different interpretations.\\

The key question is whether the two CAK parameters, $\alpha$ and $k$, are related to the properties of the wind plasma or the star's radiation field. Are they characterizing the radiation of the star (method A) or the capability of the wind material to react to radiation (method B)? In mathematical terms: 
\begin{align}
&\mathrm{method\;A}\qquad g_{rad}^l=f(\vec{u},L_{\star,1},R_{\star,1},r_1,\alpha_1,k_1)+f(\vec{u},L_{\star,2},R_{\star,2},r_2,\alpha_2,k_2)\label{mA}\\
&\mathrm{method\;B}\qquad g_{rad}^l=
	\begin{cases}
	f(\vec{u},L_{\star,1},R_{\star,1},r_1,\alpha_1,k_1)+f(\vec{u},L_{\star,2},R_{\star,2},r_2,\alpha_1,k_1) \qquad &\mbox{if in wind 1}\\
	f(\vec{u},L_{\star,1},R_{\star,1},r_1,\alpha_2,k_2)+f(\vec{u},L_{\star,2},R_{\star,2},r_2,\alpha_2,k_2) \qquad &\mbox{if in wind 2}
	\end{cases}\label{mB}
\end{align}

Both methods yield roughly the same result for CWB systems in which the two stars and the location of the WCR are far apart. However, the differences are significant for systems where this is not the case (e.g., $\eta$ Carinae during periastron, and $\gamma^2$ Velorum during its entire orbit).\\

The CAK approximation has initially been devised to describe the line acceleration of a single star and thus is to be used with great caution in the case of CWB systems. Whereas the physical interpretation of the parameter $\alpha$ is clear -- it represents the ratio of optical thin and optical thick lines in the wind plasma \citep{Cranmer1995} --, the meaning of the parameter $k$ is rather elusive, as it depends on the wind's thermal temperature whilst determining the strength of the line acceleration component \citep{Pittard2009}. In general, the two CAK parameters represent a fit to the total line driving force depending on the properties of the ions in the wind as well as on the radiation field of the star.  Accordingly, both method A and B constitute borderline cases of the problem, which, depending on the actual composition of a binary systems, should be applicable to a varying degree at the same time.\\

Past numerical simulations of CWB system have predominately used method A \citep[e.g.,][]{Pittard2009, Parkin2011, Madura2013, Reitberger2014}, whereas theoretical works on radiative inhibition and braking generally apply method B \citep[e.g.,][]{Stevens1994, Owocki1995, Gayley1997}. The latter (also known as ``strong coupling'') is slightly more complicated and costly to implement in 3D hydrodynamical simulations, as it requires the introduction of numerical means that allow to discriminate between the two wind components.\\

 The use of method A (also known as ``weak coupling'') in studies on the $\eta$ Carinae system \citep[e.g.,][]{Parkin2011} is also understandably motivated: if method B were used for this system, the turbulent WCR close to the periastron passage would be a factor of $\sim$2 closer to the surface of the WR star, thus making it even more complicated to simulate than it already is. 
This problem, however, is specific only to the $\eta$ Carinae system and its extreme values of bolometric luminosity and mass-loss rate of its primary component. For other CWB systems it is usually method B that leads to a physically more meaningful representation of the wind and, as we will show for the case of $\gamma^2$ Velorum, to improved agreement with observations linked to the shape of the WCR.\\

\subsection{Magnetohydrodynamics}
Following our extension of CWB simulations from HD to MHD \citep[as detailed in][]{Kissmann2016}, we no longer rely on any magnetic field approximation as in previous studies. In computing the synchrotron losses for the electrons accelerated at the shock we previously approximated the magnetic field following \citet{Usov1992}. Now, we use the three dimensional magnetic field components as determined by the MHD simulations. The additional information of the local direction of the magnetic field makes it possible to consider the contributions of both stars, not just the one with the dominant field. \\

As the surface magnetic field of high mass stars remains poorly constrained and as this study is not focused on the significant distortions that the presence of strong stellar dipole fields ($B_\mathrm{surface}\gtrsim 10^{-2} \mathrm{T}$) have on the structure of the WCR, we choose a moderate field strength of 10$^{-3}$ T for each star. Such a field does not produce a significant effect on the density and velocity distribution of the wind plasma. However, due to its strong influence via synchrotron losses, it still has a significant effect on the electron population. As we will show in Section \ref{results} electrons can reach higher energies along the magnetic equator because of the lower field strength and therefore lower synchrotron losses. Along the stellar dipole of the magnetic field, the losses are higher. \\

The surface magnetic field strength of the two stars are important free parameters in determining the shape of and conditions at the WCR, thus influencing not only the MHD properties of the system, but also the arising particle populations and nonthermal emission components. Whereas a choice of $B_\mathrm{surface}=10^{-3}$~T differs only insignificantly from the case of no surface magnetic field at all, higher surface magnetic field strengths lead to severe distortions in the WCR which becomes narrower, more turbulent, develops stronger curvature and also feature a nose-like structure as discussed in \citet{Kissmann2016}. Accordingly, other free parameters as the diffusion coefficient and the injection rate of high-energy protons (as discussed and identified below) have to be chosen differently for higher magnetic field strengths in order to find agreement between modeling and observations.

\subsection{Particle Acceleration}
 
In our earlier studies \citep{Reitberger2014,Reitberger2014b} the computation of diffusive shock acceleration (DSA) of charged particles at the shock fronts was done via the energy-gain rate
\begin{equation}
\dot{E}_\mathrm{DSA}=\frac{c_r-1}{3c_r\kappa}V_s^2E
\label{DSA}
\end{equation}
where $c_r$ is the compression ratio, $\kappa$ the energy-independent diffusion coefficient, $V_s$ the shock velocity and $E$ the energy of the particle. This method required the following steps: 1. the identification of ``acceleration cells'' at the shock front and determination of shock orientation therein, 2. the computation of $V_s$ as the velocity component of the wind perpendicular to the shock front,
3. the computation of $c_r$ via interpolation of the density structure perpendicular to the shock front, and
4. solving Eq. \ref{DSA} for every acceleration cell.
Steps 1 to 4 become unnecessary when using our new method to describe particle acceleration. In Appendix~A, we show that Eq. \ref{DSA} corresponds to the expression for the adiabatic cooling as used in the transport equation.
\begin{equation}
\dot{E}_\mathrm{adiab}=-\frac{E}{3}\nabla\cdot \vec{u}
\label{adiab}
\end{equation}
At the shock front, where the divergence of the velocity is negative, this term 
describes the acceleration of the particles. \\

This new method greatly simplifies the code and at the same time allows for computation of particle acceleration for turbulent shock fronts where the exact location of the acceleration region and the compression ratio are difficult to define.

\subsection{Diffusion}
Whereas maximum electron energies are governed by synchrotron and inverse Compton losses, which usually suppress the acceleration at higher energies, the maximum energies of the protons are governed by the diffusion and convection of the particles. An approximation often used is Bohm diffusion, which imposes a cutoff at maximum energies where the protons leave the shock fronts before being further accelerated. Such an approximation is no longer necessary if an energy-dependent diffusion coefficient is used \citep{Kirk1998}. \\

In earlier simulations \citep{Reitberger2014,Reitberger2014b,Reimer2006} the effects of spatial diffusion were approximated via a fundamental loss time in the transport equation -- similar to the treatment in a leaky-box model. Now, the transport equation includes a classical diffusion term $D(E)\nabla^2j$. The full equation (further motivated in Appendix 1) is
\begin{equation}
\label{D_eq}
\frac{\partial j}{\partial t}-\underset{(1)}{D(E)\nabla^2j}+\underset{(2)}{\nabla\cdot(\vec{u}j)}
-\underset{(3)}{\frac{\partial}{\partial E}\left[\left(\frac{E}{3}\nabla\cdot\vec{u}+\dot{E}_\mathrm{loss}\right)j(E)\right]}
=\underset{(4)}{Q_0\delta(E-E_0)}
\end{equation}
where $j$ is the differential number density of particles at energy $E$ at position $\vec{r}$, $\vec{u}$ is the velocity of the wind plasma and $Q_0$ is the injection rate of particles at energy $E_0$. The term $\dot{E}_\mathrm{loss}$ includes losses by inverse Compton, synchrotron and bremsstrahlung emission, Coulomb cooling, as well as losses by nucleon-nucleon collisions \citep[details in ][]{Reitberger2014}.
Whereas the spatial convection term (2) is solved along with the MHD equations via the treatment of high-energy particles as advected scalar fields, the diffusion term (1) and the energy loss and gain term (3) are solved in separate routines, both of which are capable of sub-cycling if a high diffusion coefficient $D(E)$ (or a very high energy loss rate) should require it.\\

As term (1) in equation \ref{D_eq} indicates, diffusion is treated isotropically. 
The modifications allow now to consider an energy-dependent diffusion coefficient of the form $D(E)=D_0\left(\frac{E}{1\;\mathrm{ MeV}}\right)^\delta$. The exponent $\delta$ can be set to $\delta=0.3$ for a Kolmogorov type spectrum (recommended for 3D~MHD simulations) and to $\delta=0.5$ for a Kraichnan type turbulence spectrum \citep{Strong2007}. We consistently use $\delta=0.3$ in this study (although it remains a free parameter in principle). The diffusion coefficient $D_0$ remains a free parameter. \\

In Section \ref{results} we study the effects of changing the values of the parameters $\delta$ and $D_0$ on the resulting particle spectra. Spatial diffusion as specified above determines the cutoff in the proton spectrum in most cases. \\ 

\subsection{Radiative cooling}
The hot shocked gas of the WCR is expected and observed to be a strong thermal X-ray emitter. In our model, the significant energy loss of the shocked plasma due to its thermal emission is accounted for by the radiative cooling term $\Lambda$ appearing in the energy equation \citep[see ][]{Reitberger2014}. As in the earlier versions of our code, our cooling term is based on the work of \cite{Schure2009}. However, our previous implementation of radiative cooling was only valid in the case of a wind that predominantly consists of fully ionized hydrogen ($\mu_i=1, \mu_e = 1$). Whereas this is safe to assume for OB type stars, WR stars demand a different assumption. For the corrected cooling terms considering metallic wind composition we find
\begin{equation}
\Lambda\rightarrow\frac{1}{\mu_e}\frac{1}{\mu_i}\Lambda
\end{equation}
where $\mu_i$ and $\mu_e$ depend on whether the location for which the cooling term is computed is situated within the O wind or the WR wind. To discriminate between the two wind components we make use of the same numerical means that were used in regard to the radiative wind acceleration.
Assuming the WR star to consist of fully ionized pure He ($\mu_i=4, \mu_e = 2$), a disregard of this correction due to composition would overestimate $\Lambda$ by a factor of 8. This leads to over-efficient cooling, resulting in a very dense and turbulent WCR. Once the above correction is applied we find a less turbulent WCR. 

\section{Results} 
\label{results}

\subsection{Parameters}
\label{par}
Stellar and stellar wind parameters used in our model of $\gamma^2$~Velorum along with the corresponding references are given in Table \ref{params}. We used the most up-to-date and best-constrained values found in literature, mostly relying on the fundamental parameter determination by \cite{North2007}. The CAK-parameters $\alpha$ and $k$ where fitted to obtain the desired wind parameters in a 1D simulation that is consequently adapted to 3D. Details of this procedure are found in \cite{Kissmann2016}. For the WR star we decided to run simulations for both values of the mass-loss rate that have been determined either by polarimetric measurements ($\dot{M}$=$8\times10^{-5}$ M$_{\odot}$ yr$^{-1}$) or via radio-continuum observations ($\dot{M}$=$3\times10^{-6}$ M$_{\odot}$ yr$^{-1}$). The figures and values presented in this publication show the results determined with the latter, which we deem more likely due to reasons given below. Note that in \cite{Reitberger2017} we used the former mass-loss rate which led to different results. \\

The orbital parameters are shown in Table \ref{orbital}. Our choice reflects the so far best constrained distance determination and reliable values for stellar separation and eccentricity. The inclination $i$ of the orbital plane of the $\gamma^2$~Velorum binary system and its argument of apastron $\omega_\mathrm{WR}$ are well constrained. \cite{Schmutz1997} find $i=65^\circ \pm 8^\circ$ and $\omega_\mathrm{WR}=68^\circ \pm 4^\circ$. The exact definition of the angle $\omega_\mathrm{WR}$ becomes clear by the consideration that if the inclination were 90$^\circ$ (edge-on view), the WR star would eclipse the O star at the apastron passage if $\omega_\mathrm{WR}=0$, and $\sim9$~days after apastron if $\omega_\mathrm{WR}=68^\circ$. Thus, the meaning of the angle $\omega_\mathrm{WR}$ \citep[as in][]{Schmutz1997} is consistent with the definition of the angle $\Phi$ in \cite{Reitberger2014b}. \\

All other parameters relevant for particle acceleration or nonthermal $\gamma$-ray emission are either the same as in \cite{Reitberger2014b,Reitberger2014} or individually discussed in the sections below.


\begin{deluxetable}{cccccccccc}
\tablecaption{Stellar and stellar wind parameters of $\gamma^2$~Velorum used in this study, taken from [1] \cite{deMarco2000}, [2] \cite{North2007}, [3] \cite{Schild2004}  \label{params}}
\tablecolumns{9}
\phantom{$\arcmin$}
\tablehead{\colhead{Star} & \colhead{$M_\ast$} & \colhead{$R_\ast$} & \colhead{$T_\ast$} & \colhead{$L_\ast$ } & \colhead{$\dot{M}$} & \colhead{$v_\infty$} & \colhead{$\alpha$} & \colhead{$k$} & $B_\mathrm{surface}$ \\ 
\colhead{} & \colhead{($M_{\odot}$)} & \colhead{($R_{\odot}$)} & \colhead{(K)} & \colhead{($L_{\odot}$)} & \colhead{($M_{\odot}$ yr$^{-1}$)} & \colhead{(km s$^{-1}$)} & \colhead{-} & \colhead{-} & \colhead{T}} 
\startdata
O7.5 &   28.5 [1,2]  & 17 [2]  &  35000 [1]  & 2.8$\times10^{5}$ [2]  & $1.78\times10^{-7}$ [1] & 2500 [1] & 0.613 & 0.055 & 10$^{-2}$ \\
\hline
\multirow{2}{*}{WC8}  & \multirow{2}{*}{9 [2]} & \multirow{2}{*}{6 [2]} & \multirow{2}{*}{56000 [2]} & \multirow{2}{*}{$1.7\times 10^{5}$ [2]} & $8\times10^{-6}$ [2,3]  & \multirow{2}{*}{1450 [2,3]}& \multirow{2}{*}{0.526} & 0.747 & \multirow{2}{*}{10$^{-2}$} \\
&  &  & & & $3\times10^{-5}$ [1,2]  &   &  & 1.498 & \\
\enddata
\end{deluxetable}
\begin{deluxetable}{cccccc}
\tablecaption{Orbital parameters used in this study, taken from [1] \cite{North2007}, [2] \cite{Schmutz1997}  \label{orbital}}
\tablecolumns{9}
\phantom{$\arcmin$}
\tablehead{\colhead{distance $d$} & \colhead{semi-major axis $a$} & \colhead{period $P$} & \colhead{eccentricity $e$} & \colhead{inclination $i$} & \colhead{$\omega_{WR}=\Phi$ } \\ 
\colhead{(pc)} & \colhead{($R_\odot$)} & \colhead{(d)} & \colhead{-} & \colhead{($^\circ$)} & \colhead{($^\circ$)}} 
\startdata
336 [1] &   258 [1]  & 78.53 [2]  &  0.334 [2]  & 65 [2] & 68 [2] \\
\enddata
\end{deluxetable}

\subsection{MHD results}
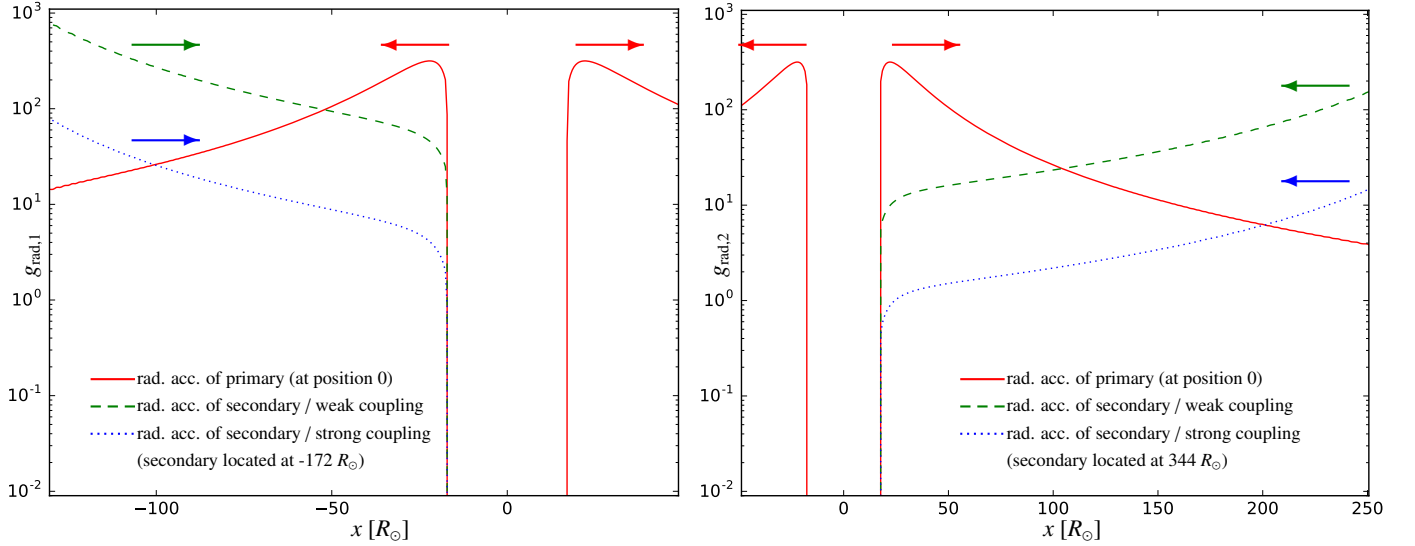
\begin{figure}[t!]

	\linethickness{0.3mm}
	\setlength{\unitlength}{0.0005\textwidth}
	
	\begin{picture}(1000,800)(0,0)
	\put(0,0){\includegraphics[width=984\unitlength]{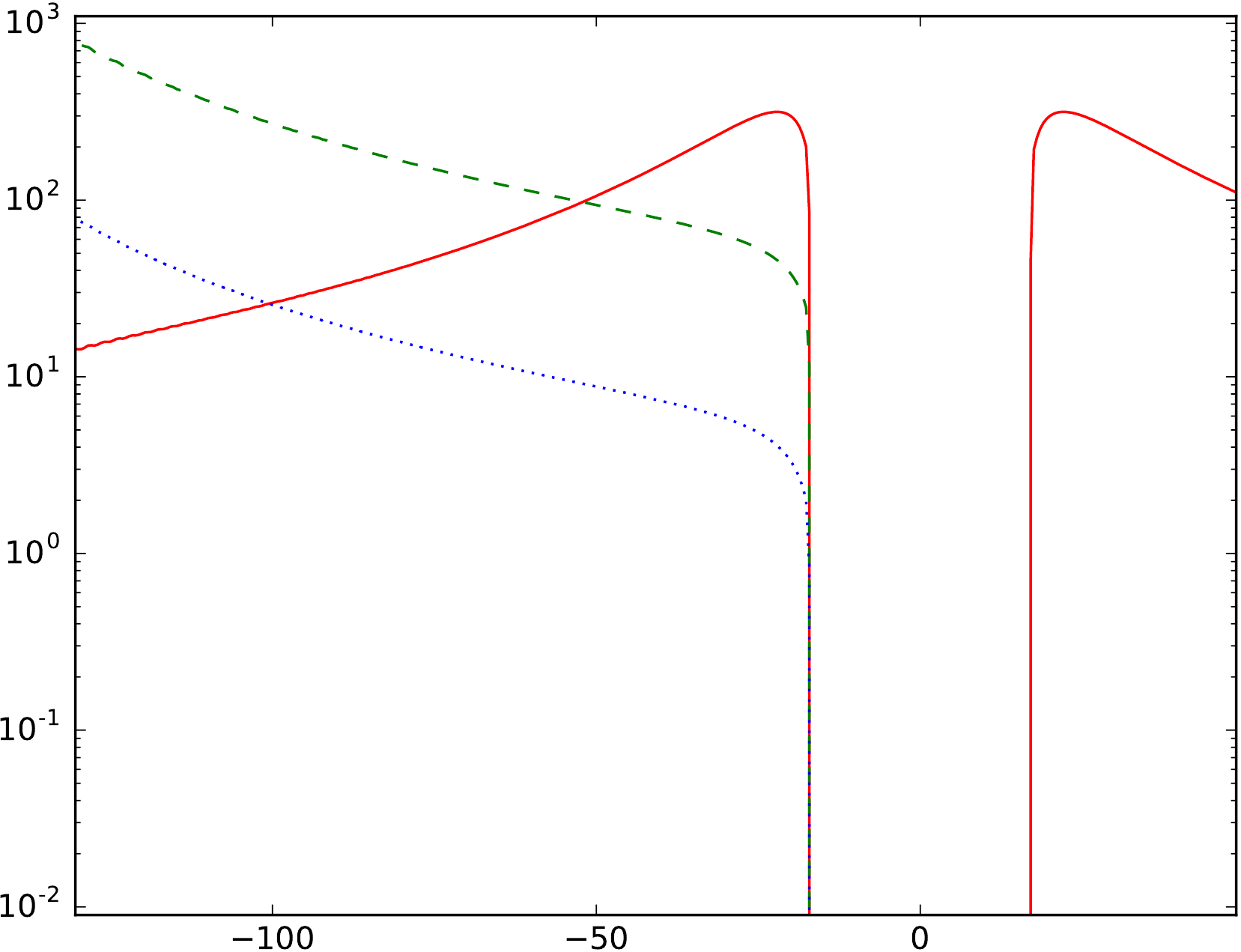}}
	\put(25,350){\rotatebox{90}{$g_\mathrm{rad,1} $}}
	\put(500,-30){$x$ [$R_{\sun}$]}
	\thicklines \put(120,200){\color{red}\line(1,0){65}}
	\thicklines \multiput(120,160)(25,0){3}{\color{green!50!black}\line(1,0){15}}
	\thicklines \multiput(120,120)(10,0){7}{\color{blue}\line(1,0){3}}
	\put(188,193){\scriptsize rad. acc. of primary (at position 0)}
	\put(188,153){\scriptsize rad. acc. of secondary / weak coupling}
	\put(188,113){\scriptsize rad. acc. of secondary / strong coupling}
	\put(188,73){\scriptsize (secondary located at -172 $R_{\sun}$)}
	\put(645,690){\color{red}\vector(-1,0){100}}
	\put(830,690){\color{red}\vector(1,0){100}}
	\put(180,690){\color{green!50!black}\vector(1,0){100}}
	\put(180,550){\color{blue}\vector(1,0){100}}
	\end{picture}
	\begin{picture}(1000,800)(0,0)
	\put(0,0){\includegraphics[width=1000\unitlength]{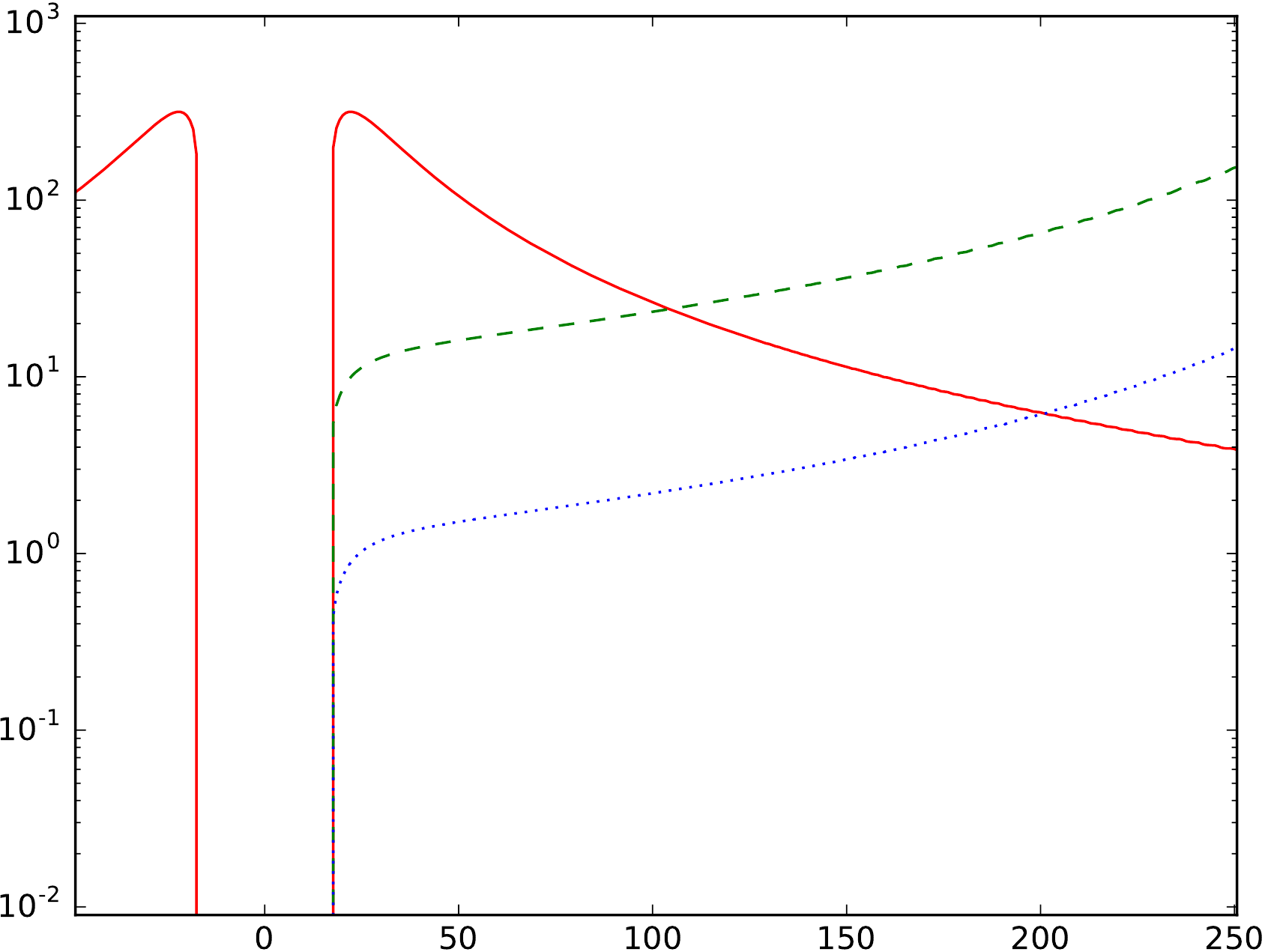}}
	\put(20,350){\rotatebox{90}{$g_\mathrm{rad,2} $}}
	\put(500,-30){$x$ [$R_{\sun}$]}
	\thicklines \put(380,200){\color{red}\line(1,0){65}}
	\thicklines \multiput(380,160)(25,0){3}{\color{green!50!black}\line(1,0){15}}
	\thicklines \multiput(380,120)(10,0){7}{\color{blue}\line(1,0){3}}
	\put(448,193){\scriptsize rad. acc. of primary (at position 0)}
	\put(448,153){\scriptsize rad. acc. of secondary / weak coupling}
	\put(448,113){\scriptsize rad. acc. of secondary / strong coupling}
	\put(448,73){\scriptsize (secondary located at 344 $R_{\sun}$)}
	\put(155,690){\color{red}\vector(-1,0){100}}
	\put(280,690){\color{red}\vector(1,0){100}}
	\put(950,630){\color{green!50!black}\vector(-1,0){100}}
	\put(950,490){\color{blue}\vector(-1,0){100}}
	\end{picture}
				\caption{Absolutes of the radiative acceleration in the primary wind close to the primary star (located at position 0) for periastron (left) and apastron (right) configuration. The accleration felt by the primary wind due to the radiation of the secondary star is indicated for the case of weak (green dashed line) and strong (blue dotted line) coupling. Arrows indicate the direction of the various acceleration components either towards or away from the primary. The intersections mark the points of balance where the radiative components are equal. 
				\label{CAKcoupling} 
		 }
\end{figure}

In a short-period binary system like $\gamma^2$~Velorum, the shape and plasma conditions of the WCR strongly depend on the choice of weak or strong coupling. 
Fig. \ref{CAKcoupling} shows the possible components of the total line acceleration $g^l_\mathrm{rad}$ as in Equations \ref{mA} and \ref{mB} in the primary wind along the line of centers connecting the two stars. The acceleration acting on the primary wind by the primary star's radiation field  $f(\vec{u},L_{\star,1},R_{\star,1},r_1,\alpha_1,k_1)$ is shown in solid red, the acceleration acting on the primary wind by the secondary star's radiation field is shown in dashed green for the case of weak coupling ($f(\vec{u},L_{\star,2},R_{\star,2},r_2,\alpha_2,k_2)$, method A) and in dotted blue for the case of strong coupling ($f(\vec{u},L_{\star,2},R_{\star,2},r_2,\alpha_1,k_1)$, method B).\\

The computation of $g^l_\mathrm{rad}$ for all three cases is based on the primary wind's initial velocity profile neglecting the presence of the secondary. Fig. \ref{CAKcoupling} shows the system at a state before the influence of the secondary acts. It indicates that -- even without the influence of the ram pressure of the secondary wind -- the primary wind will be pushed back towards the primary star merely by the acceleration of the secondary's radiation field. For the case of weak coupling the distance where the two oppositely oriented acceleration components cancel out is merely $\sim 35 \mathrm{R}_\odot$ above the stellar surface for periastron and $\sim$85 $\mathrm{R}_\odot $ for apastron. After including the presence of the secondary's wind (and not just the secondary star's radiative influence) the wind's ram pressure will push the WCR even closer towards the O~star. For the case of strong coupling the distance of equal radiative acceleration components is $\sim 83 \mathrm{R}_\odot$ from the stellar surface for periastron, and  $\sim 180 \mathrm{R}_\odot$ for apastron configuration. There is more than a factor of 2 between the respective values for strong and weak coupling.\\

\begin{figure}[t!]
	\setlength{\unitlength}{0.0005\textwidth}
	\begin{picture}(2000,1700)(0,0)
	\put(0,850){\includegraphics[trim={0.0cm 0 3cm 0},clip,width=880\unitlength]{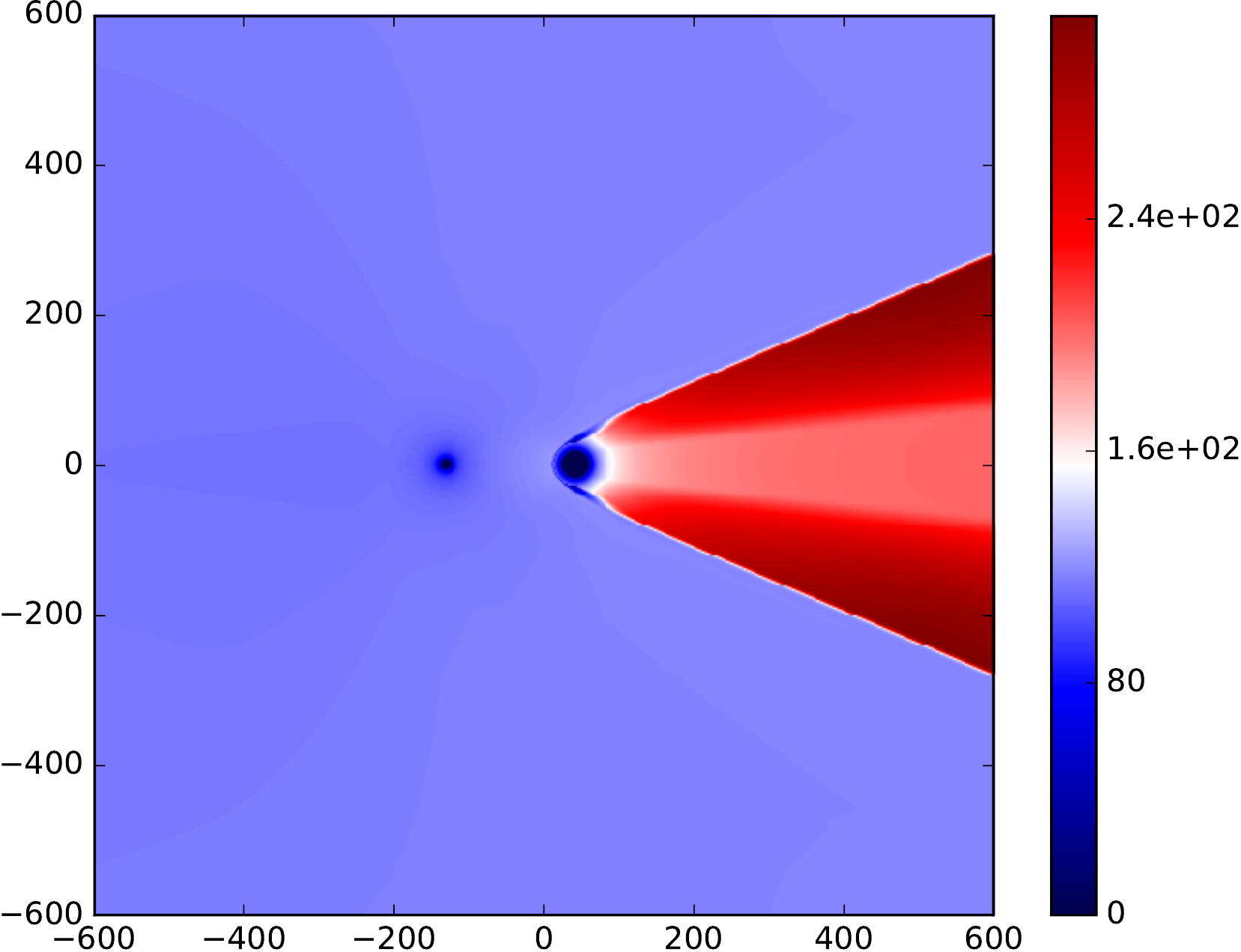}}
	\put(900,850){\includegraphics[trim={0.0cm 0 3cm 0},clip,width=880\unitlength]{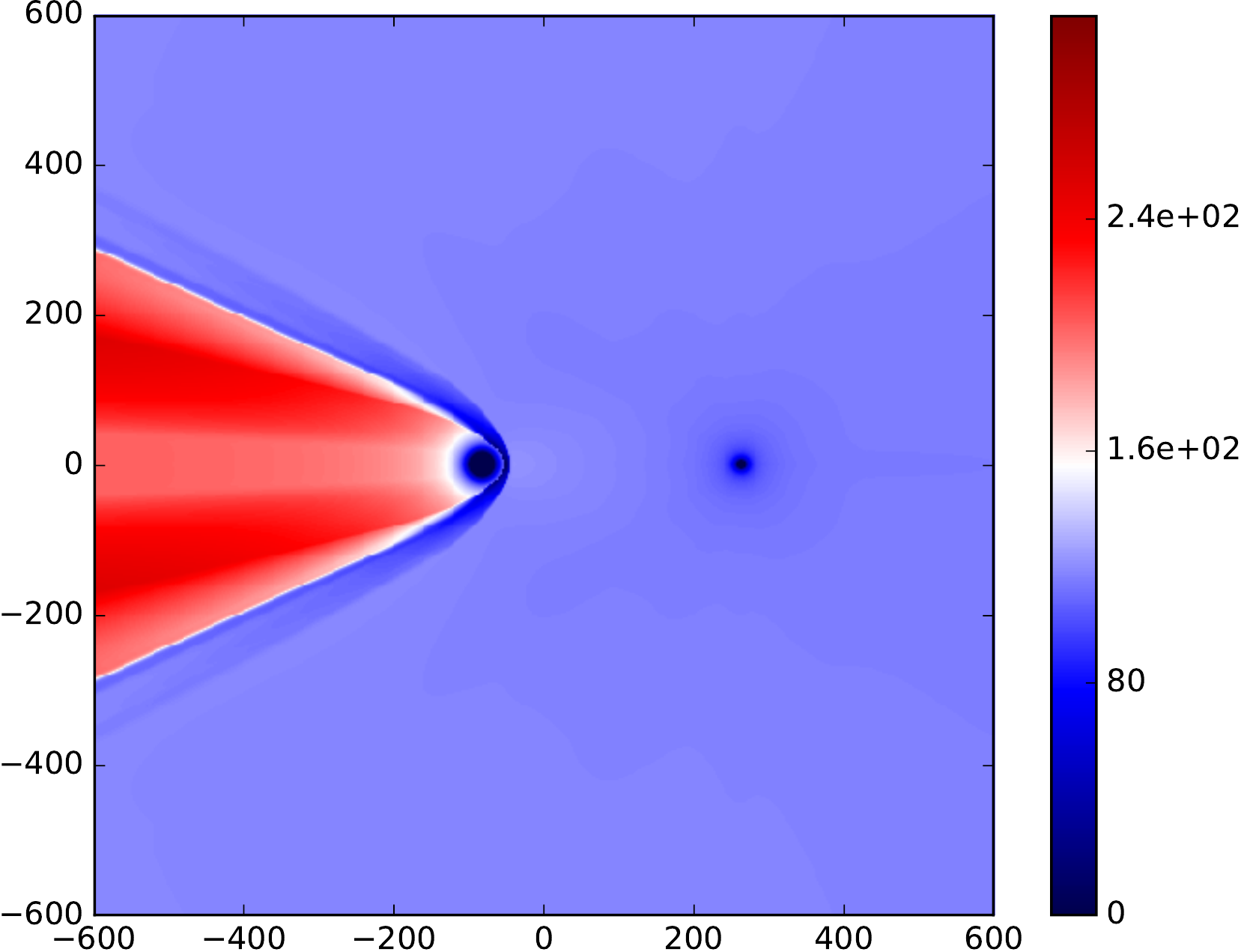}}
	\put(0,0){\includegraphics[trim={0.0cm 0 3cm 0},clip,width=880\unitlength]{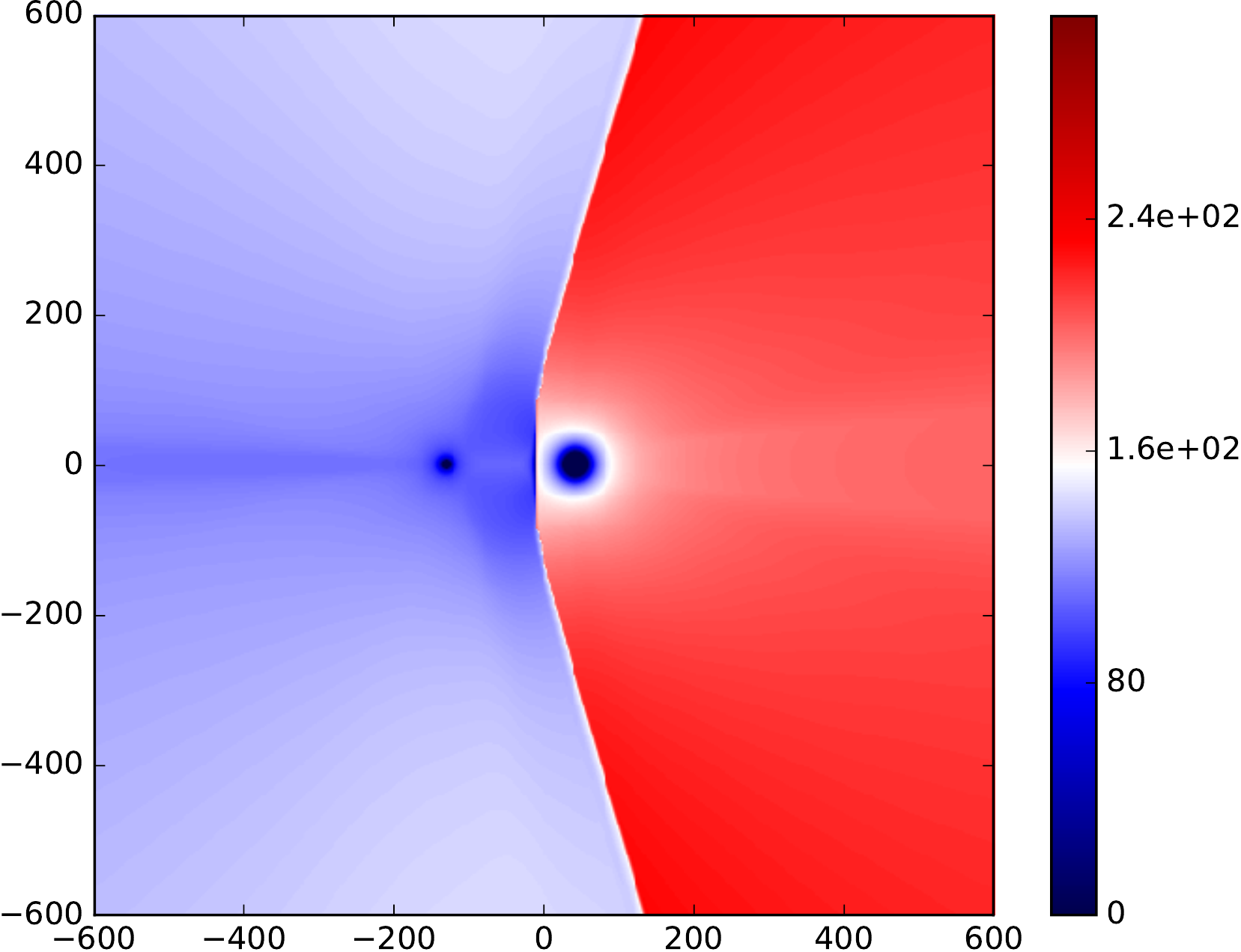}}
	\put(900,0){\includegraphics[trim={0.0cm 0 3cm 0},clip,width=880\unitlength]{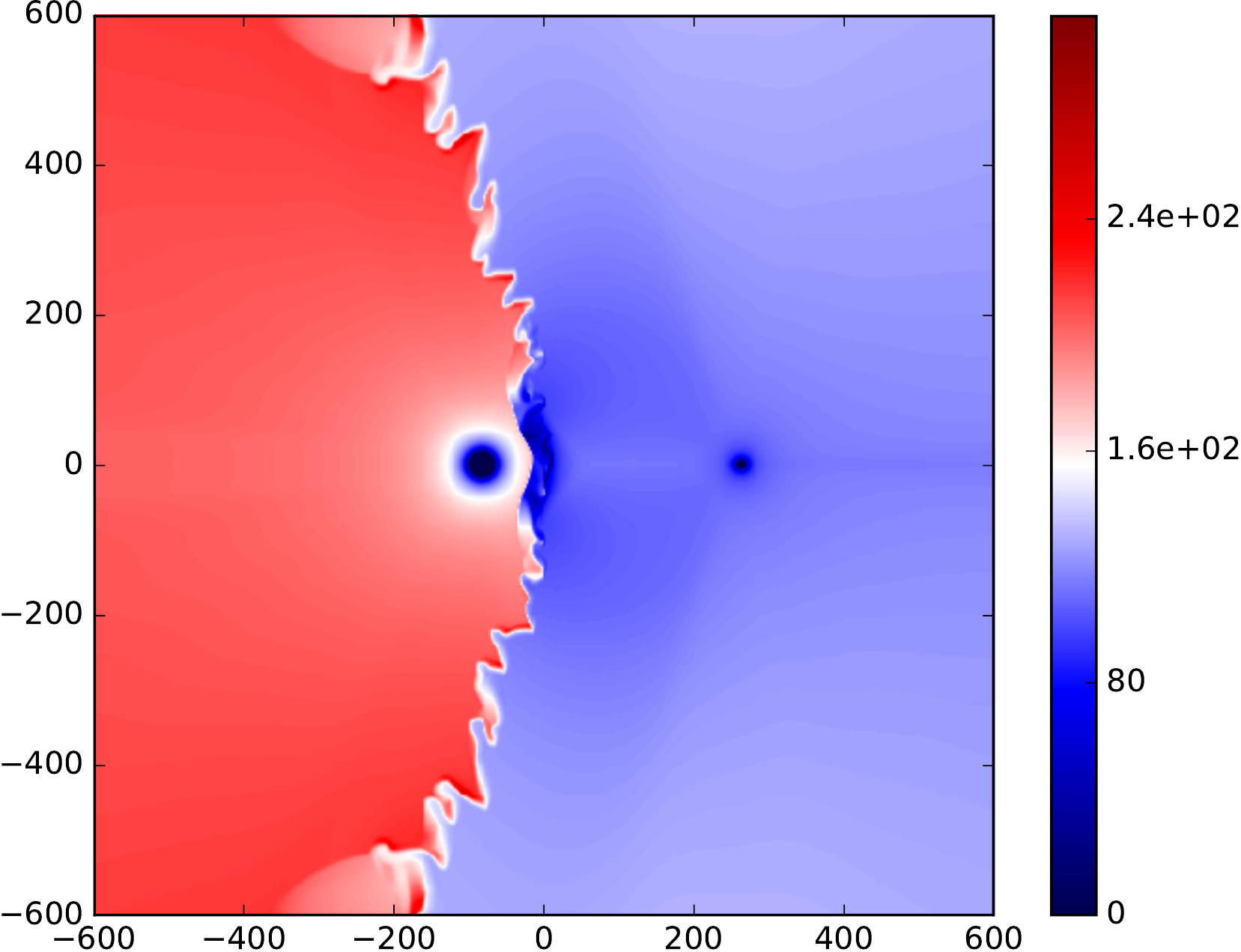}}	
	\put(1800,0){\includegraphics[trim={14.2cm 0.25cm 1.8cm 0cm},clip,height=1680\unitlength]{CAKvabs_strong_ap_512.pdf}}
	\put(1910,25){\large 0}
	\put(1910,430){\large 0.8}
	\put(1910,850){\large 1.6}
	\put(1910,1260){\large 2.4}
	\put(1780,-30){\large $10^6$ m s$^{-1}$}
	\put(500,-30){$x$ [$R_{\sun}$]}
	\put(1400,-30){$x$ [$R_{\sun}$]}
	\put(20,1240){\rotatebox{90}{$y$ [$R_{\sun}$]}}
	\put(20,390){\rotatebox{90}{$y$ [$R_{\sun}$]}}
	\put(500,450){\textcolor{yellow}{\Huge$\theta$}}
	\put(1270,450){\textcolor{yellow}{\Huge$\theta$}}
	\put(580,1280){\textcolor{yellow}{\Huge$\theta$}}
	\put(1190,1280){\textcolor{yellow}{\Huge$\theta$}}
	\put(100,1550){\textcolor{white}{\huge method A}}
	\put(100,700){\textcolor{white}{\huge method B}}
	\put(365,360){\textcolor{white}{ WR}}
	\put(365,1220){\textcolor{white}{ WR}}
	\put(485,360){\textcolor{white}{O}}
	\put(485,1220){\textcolor{white}{O}}
	\put(1300,1220){\textcolor{white}{O}}
	\put(1520,1220){\textcolor{white}{WR}}
	\put(1300,360){\textcolor{white}{O}}
	\put(1520,360){\textcolor{white}{WR}}
	\put(0,0){\begin{tikzpicture}[ link/.style    = {dashed, line width = 2pt,
                    },]
		\coordinate (A7) at (0\unitlength,0\unitlength);
		\coordinate (a) at (425\unitlength,1265\unitlength);
		\coordinate (b) at (920\unitlength,1477\unitlength);
		\coordinate (c) at (920\unitlength,1265\unitlength);
	\draw[fill=black] (A7) circle (5\unitlength);
	 \draw[link,yellow] (a) -- (b);
	 \draw[link,yellow] (a) -- (c);
			\coordinate (d) at (440\unitlength,415\unitlength);
		\coordinate (e) at (555\unitlength,827\unitlength);
		\coordinate (f) at (920\unitlength,415\unitlength);
	 \draw[link,yellow] (d) -- (e);
	 \draw[link,yellow] (d) -- (f);
	 		\coordinate (g) at (1380\unitlength,415\unitlength);
		\coordinate (h) at (1230\unitlength,827\unitlength);
		\coordinate (i) at (920\unitlength,415\unitlength);
	 \draw[link,yellow] (g) -- (h);
	 \draw[link,yellow] (g) -- (i);
	 \coordinate (j) at (1370\unitlength,1265\unitlength);
		\coordinate (k) at (920\unitlength,1477\unitlength);
		\coordinate (l) at (920\unitlength,1265\unitlength);
	 \draw[link,yellow] (j) -- (k);
	 \draw[link,yellow] (j) -- (l);
	\end{tikzpicture}}
	\end{picture}
	\caption{Absolute velocity of the wind plasma for periastron (left) and apastron (right) with weak coupling (upper panel) and strong coupling (lower panel). The approximate opening angle $\theta$ is indicated by the dashed yellow lines. \label{vabs}} 
\end{figure}

This significant difference between weak and strong coupling is shown in Fig. \ref{vabs} -- now including the presence of the secondary's wind. In both the apastron and the periastron state, weak coupling leads to a narrow shock-cone of the primary wind which is engulfed by a dominant secondary wind from the WR star. However, by using strong coupling we find a much larger opening angle. The wind collision happens also further away from the primary with strong turbulence in the apastron configuration. The blue-shaded regions of the secondary wind between the two stars show effects of radiative braking where the velocity of the secondary-wind is significantly lowered by the primary star's radiation field. The narrow blue region on the far-side of the WR star results from the effect of shadowing in this region. The primary star is obscured by the secondary. Therefore the wind plasma in the shadow of the secondary does not feel any radiative acceleration component caused by the primary. \\

The choice of weak or strong coupling not only influences the distance of the WCR from the primary star but also severely alters the opening half-angle of the shock cone: $\sim 24^\circ$ in the case of weak coupling, $\sim 72^\circ$ in the case of strong coupling. As already mentioned in Secton \ref{intro}, there is evidence from high-resolution X-ray spectroscopy of Chandra data, \citep{Henley2005} that $\gamma^2$~Velorum exhibits a large shock-cone opening half-angle of $\sim 85^\circ$. In the same study the authors remark that simulations hitherto failed to reproduce such a feature which might be due to significant radiative braking. In our simulation, including the effects of radiative braking with strong coupling of the CAK parameters, we can indeed reproduce this large shock-cone opening angle, which is most evident close to the periastron passage (see Fig. \ref{vabs}, lower panel). Consequently sufficient agreement with observations from X-ray spectroscopy can only be achieved by using strong coupling to the wind. We therefore consistently use this method in the following analysis.\\

\begin{figure}
	\setlength{\unitlength}{0.0008\textwidth}
		\begin{picture}(335,321)(0,0)
		\put(0,0){\includegraphics[height=320\unitlength,trim=0cm 0cm 2.2cm 0 cm, clip=true]{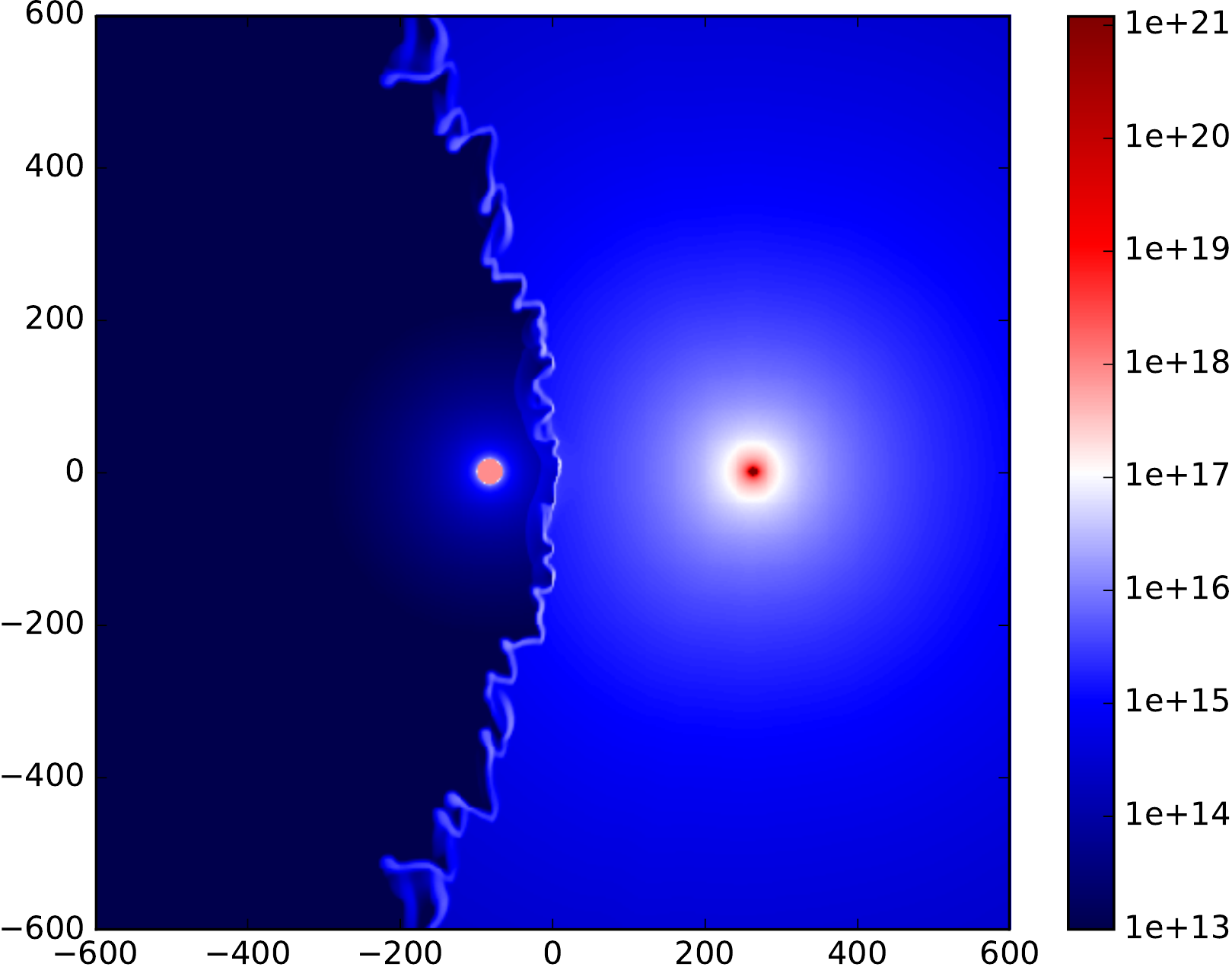}}
		\put(0,150){\rotatebox{90}{\scriptsize $z\;[\mathrm{R}_\odot] $}}
	\end{picture}	
		\begin{picture}(50,321)(0,0)
		\put(0,0){\includegraphics[height=320\unitlength,trim=14cm 0cm 0cm 0 cm, clip=true]{ap_rho.pdf}}
	\end{picture}
		\begin{picture}(335,321)(0,0)
		\put(0,0){\includegraphics[height=320\unitlength,trim=0cm 0cm 2.2cm 0 cm, clip=true]{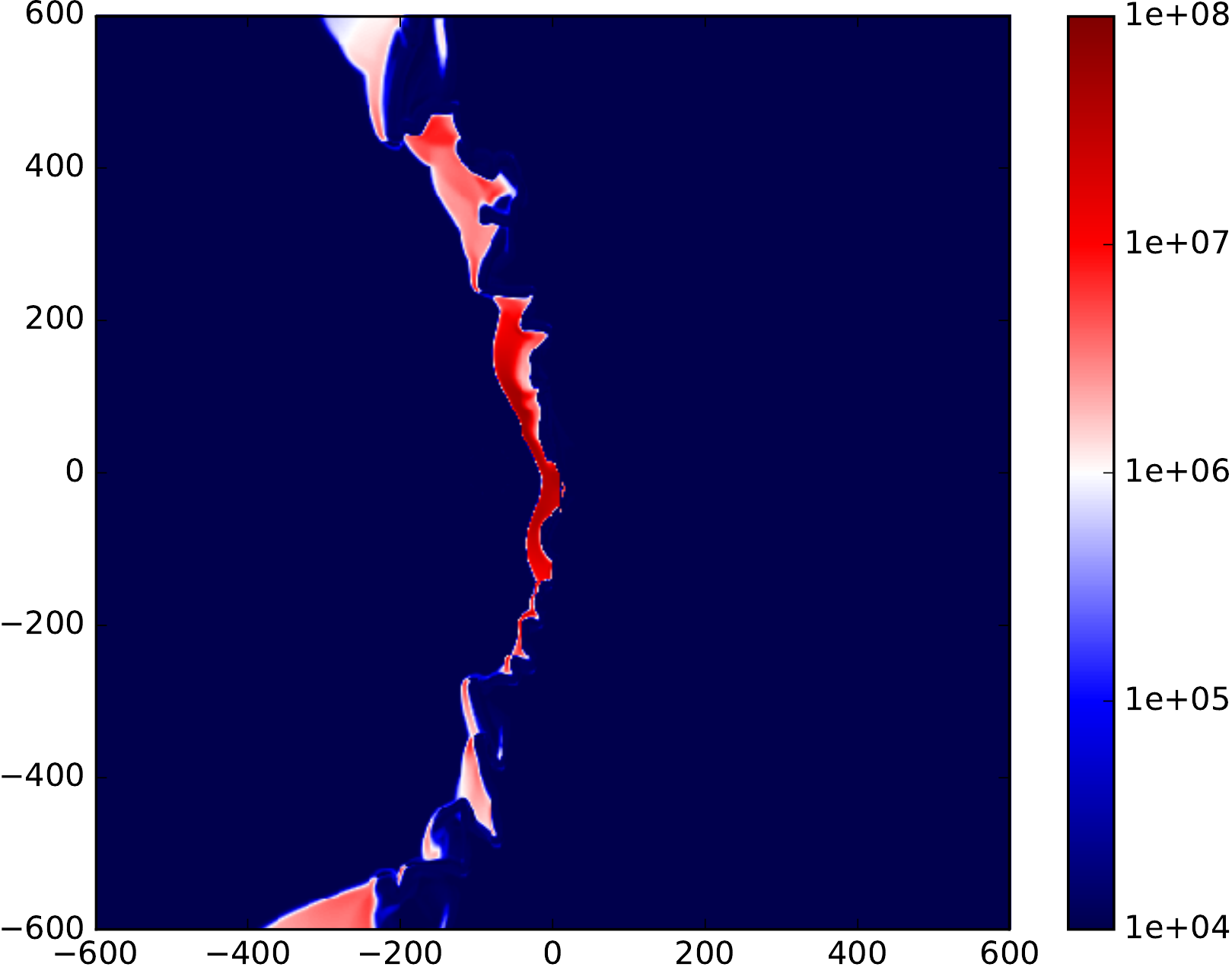}}
	\end{picture}		
		\begin{picture}(50,321)(0,0)
		\put(0,0){\includegraphics[height=320\unitlength,trim=14cm 0cm 0cm 0 cm, clip=true]{ap_Temp.pdf}}
	\end{picture}
		\begin{picture}(335,321)(0,0)
		\put(0,0){\includegraphics[height=320\unitlength,trim=0cm 0cm 2.2cm 0 cm, clip=true]{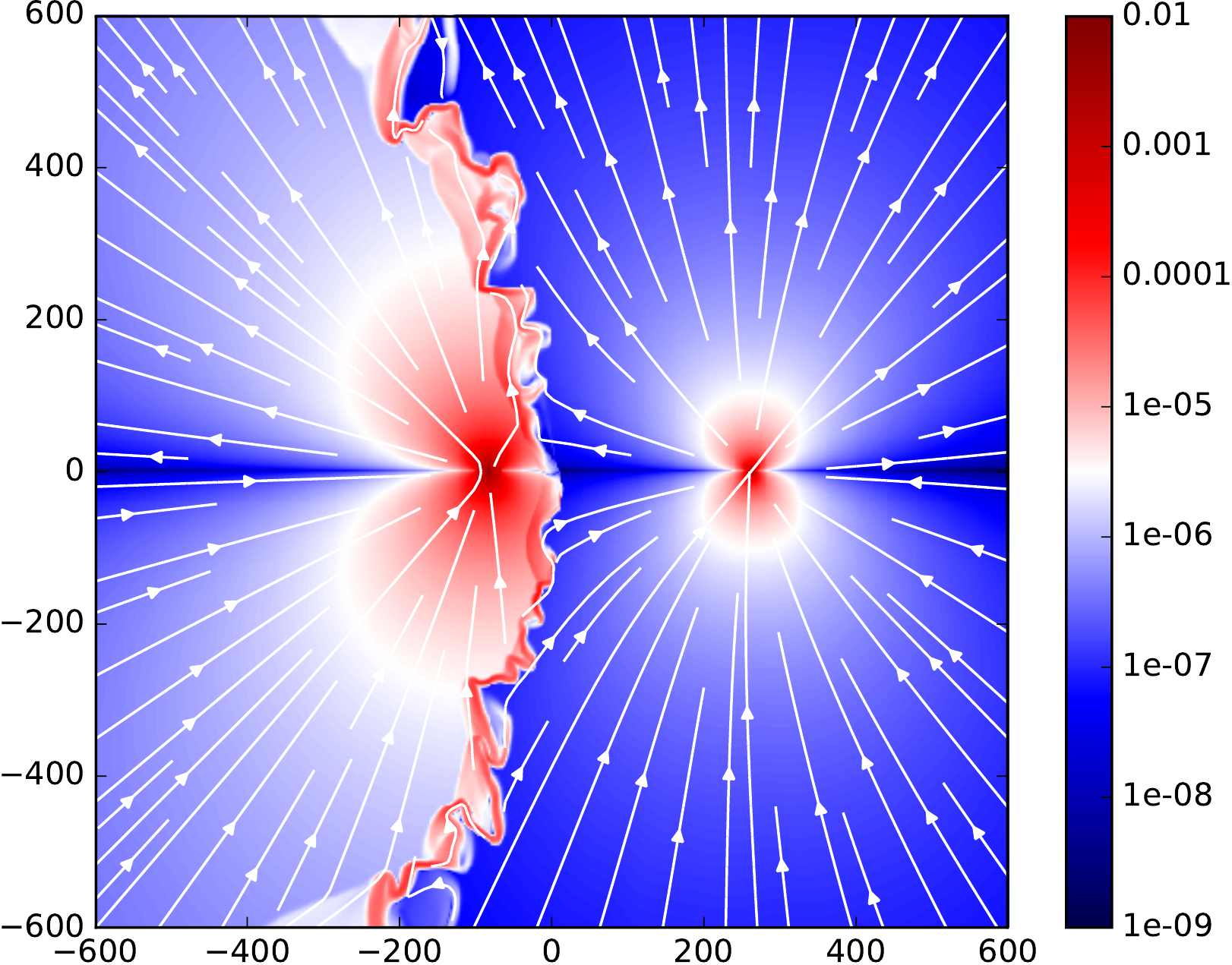}}
	\end{picture}			
		\begin{picture}(50,321)(0,0)
		\put(0,0){\includegraphics[height=320\unitlength,trim=14cm 0cm 0cm 0 cm, clip=true]{ap_B_xz.pdf}}
	\end{picture}\\
			\begin{picture}(335,321)(0,0)
		\put(0,0){\includegraphics[height=320\unitlength,trim=0cm 0cm 2.2cm 0 cm, clip=true]{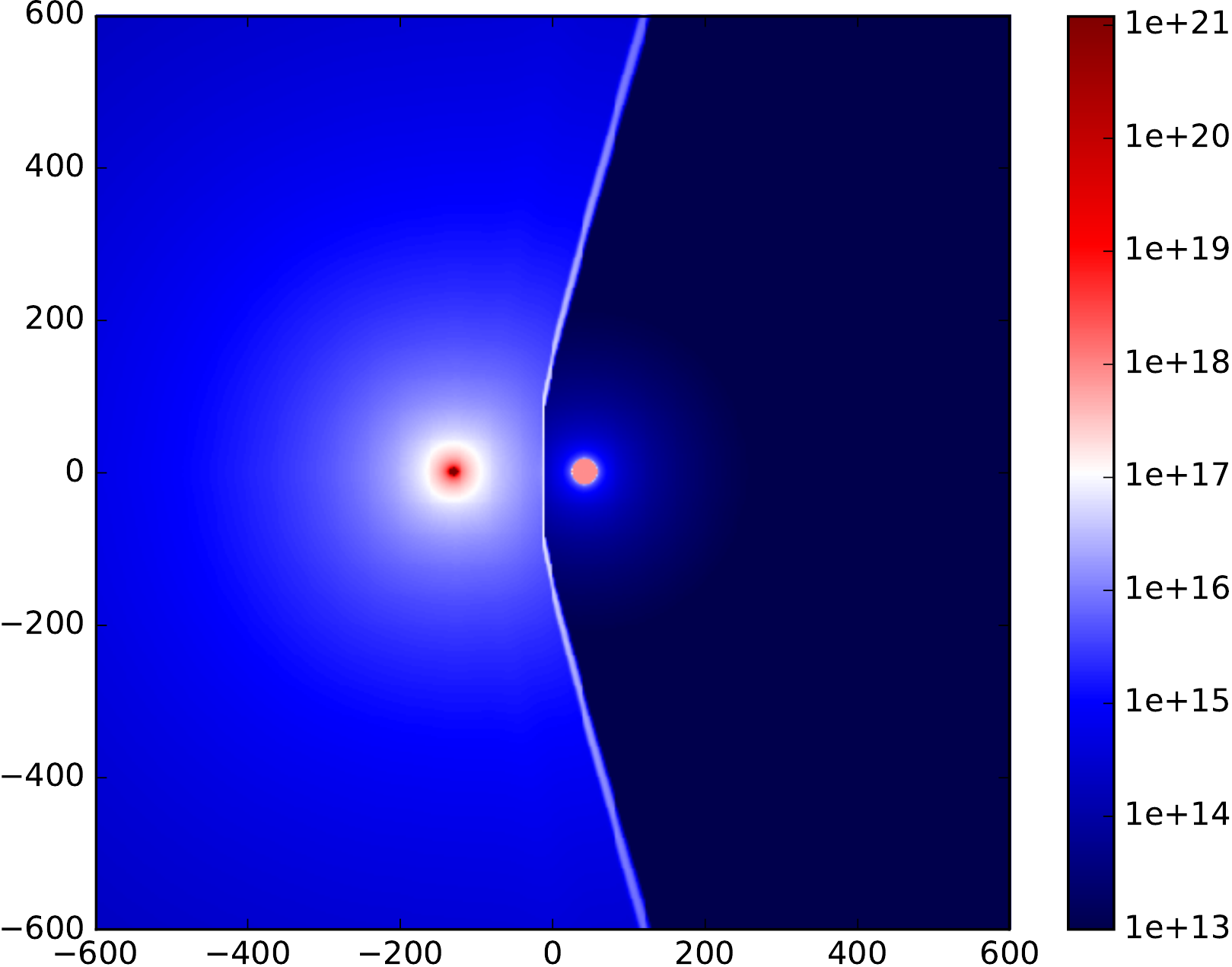}}
		\put(0,150){\rotatebox{90}{\scriptsize $z\;[\mathrm{R}_\odot] $}}
		\put(150,-15){\scriptsize $x\;[\mathrm{R}_\odot] $}
	\end{picture}	
		\begin{picture}(50,321)(0,0)
		\put(0,0){\includegraphics[height=320\unitlength,trim=14cm 0cm 0cm 0 cm, clip=true]{pe_rho.pdf}}
		\put(0,-5){\scriptsize $n \; [\mathrm{m^{-3}]}$}
	\end{picture}
		\begin{picture}(335,321)(0,0)
		\put(0,0){\includegraphics[height=320\unitlength,trim=0cm 0cm 2.2cm 0 cm, clip=true]{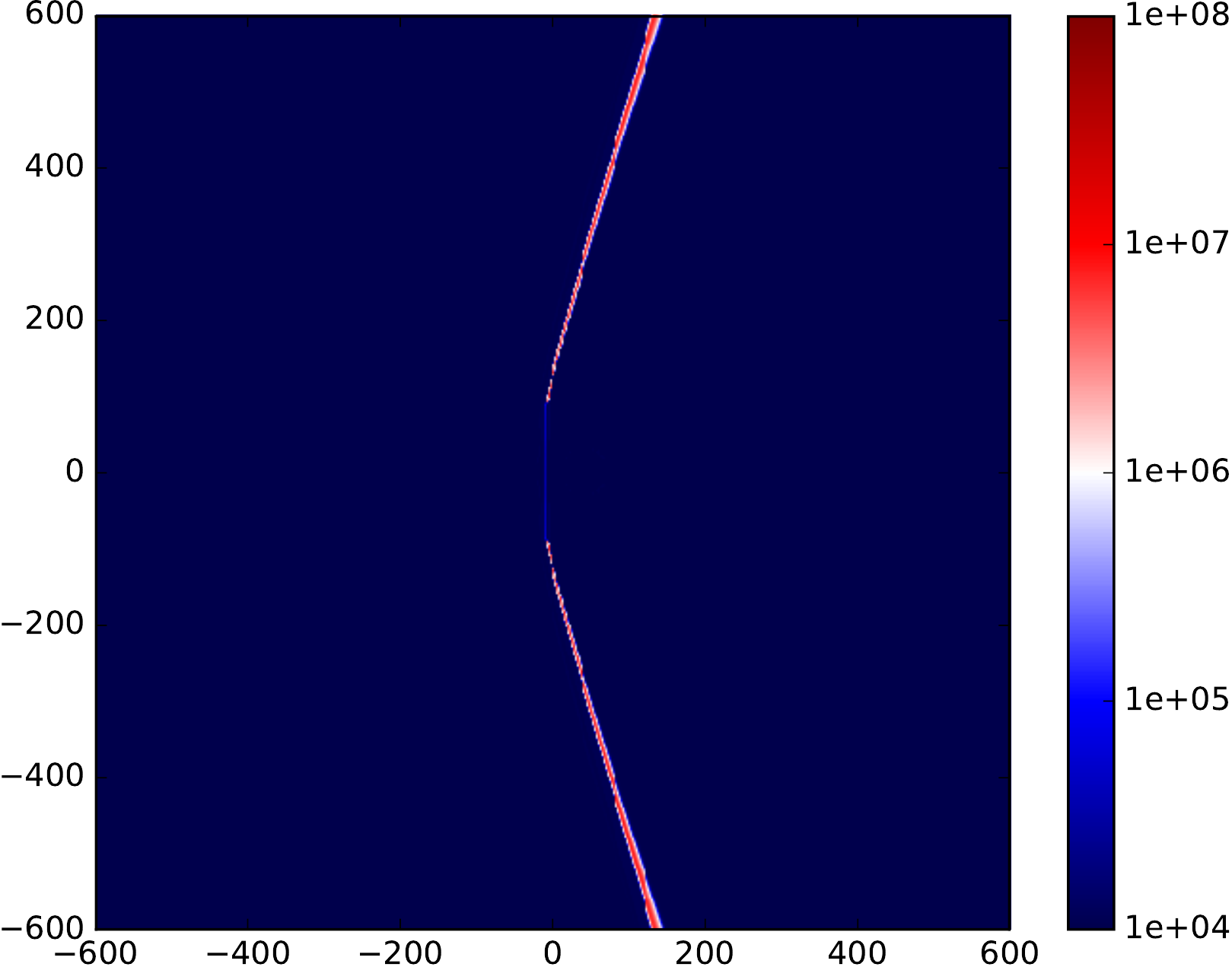}}
		\put(150,-15){\scriptsize $x\;[\mathrm{R}_\odot] $}
	\end{picture}		
		\begin{picture}(50,321)(0,0)
		\put(0,0){\includegraphics[height=320\unitlength,trim=14cm 0cm 0cm 0 cm, clip=true]{pe_Temp.pdf}}
		\put(0,-5){\scriptsize $T\;[\mathrm{K}] $}
	\end{picture}
		\begin{picture}(335,321)(0,0)
		\put(0,0){\includegraphics[height=320\unitlength,trim=0cm 0cm 2.2cm 0 cm, clip=true]{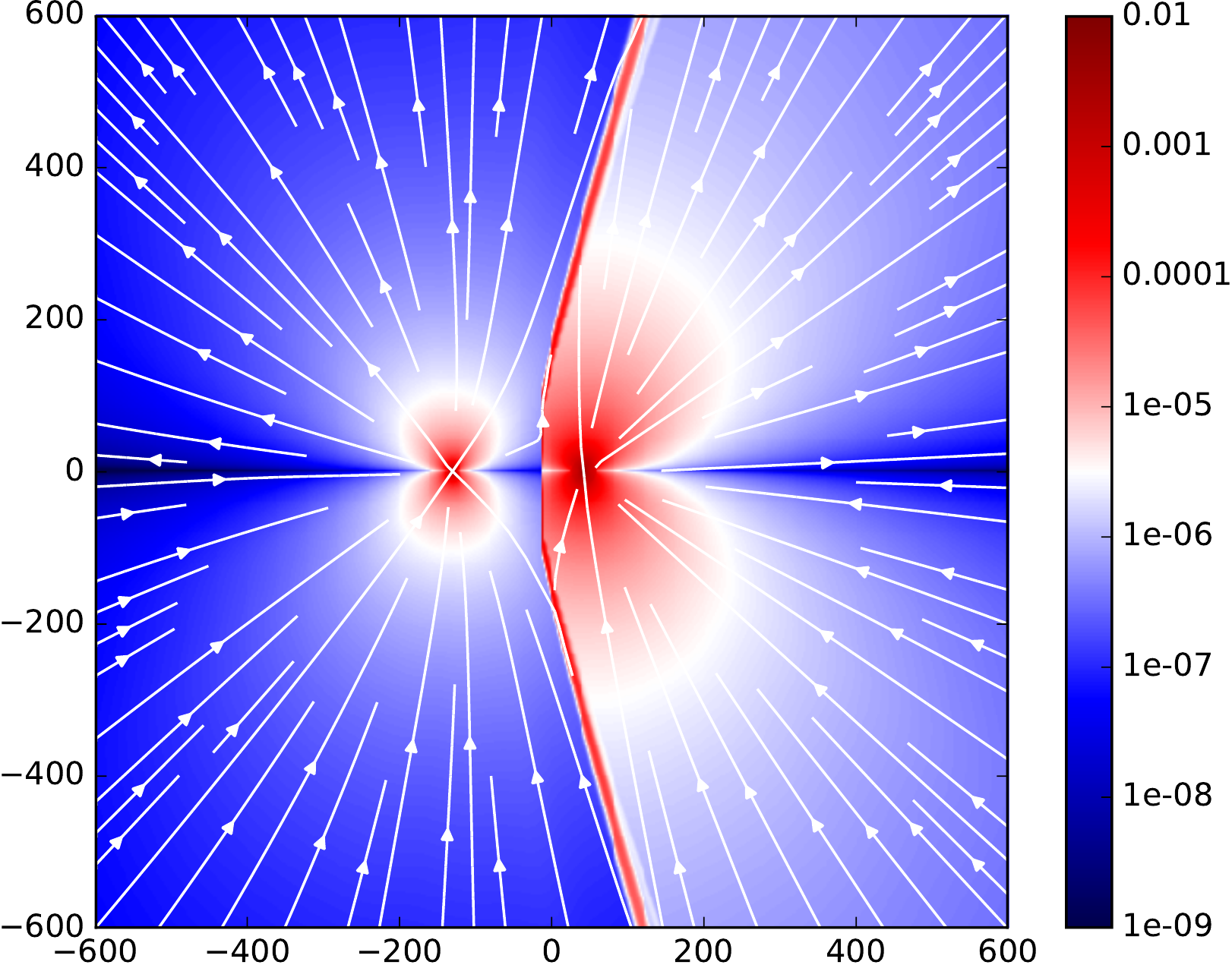}}
		\put(150,-15){\scriptsize $x\;[\mathrm{R}_\odot] $}
	\end{picture}			
		\begin{picture}(50,321)(0,0)
		\put(0,0){\includegraphics[height=320\unitlength,trim=14cm 0cm 0cm 0 cm, clip=true]{pe_B_xz.pdf}}
		\put(0,-5){\scriptsize $B\;[\mathrm{T}] $}
	\end{picture}
	
	\caption{Number density of wind particles (left), temperature (center), and magnetic field strength (right) of $\gamma^2$~Velorum in the x--z plane for apastron (top) and periastron (bottom) configuration.}
	\label{MHD1}
\end{figure}

The remaining fluid properties of density, temperature, and magnetic field strength are shown in Fig. \ref{MHD1}. 
The density of the wind plasma has a similar structure as the velocity: a thin laminar WCR for periastron and a turbulent WCR for apastron configuration. Maximum densities inside the WCR are $\sim$5$\times 10^{16}$ m$^{-3}$ at the apex of both cases.
The wind plasma reaches temperatures up to ~5$\times$10$^7$~K in the turbulent WCR for the apastron configuration. Periastron temperatures are slightly lower and reach a maximum of just below 10$^6$~K in the inner regions of the laminar WCR. The apex of the periastron region remains cool due to severe radiative cooling in the high-density environment of the thin WCR.
Regarding the magnetic field, the high field strengths of the O star's dipole strongly influences the conditions at the WCR where field strengths of 10$^{-5}$~T are reached. Electrons accelerated close to the apex of the WCR will suffer from severe synchrotron losses due to the influence of the O star's dipole field. This is also valid for periastron configuration where the more laminar WCR is also close to regions with high magnetic field strength. Whereas inverse Compton losses are clearly dominant in the x--y plane where magnetic fields are low, the order reverses in regions of higher magnetic field strengths where losses by synchrotron  emission dominate all others.

\subsection{High-energy particle results}

By solving the transport equation (along with the diffusion equation and the MHD equations including the spatial convection) for 100 logarithmically spaced energy bins between 1 MeV and 10 TeV for electrons and protons each, we obtain populations of high energy particles at every location throughout the simulated volume. Details are given in \cite{Reitberger2014,Reitberger2014b}. As for the updated treatment of diffusion we first study the effects of the two parameters $\delta$ and $D_0$ for a single grid cell only. \\

\begin{figure}[t]
\begin{center}
	\setlength{\unitlength}{0.001\textwidth}
	\begin{picture}(320,280)(0,0)
		\includegraphics[width=320\unitlength]{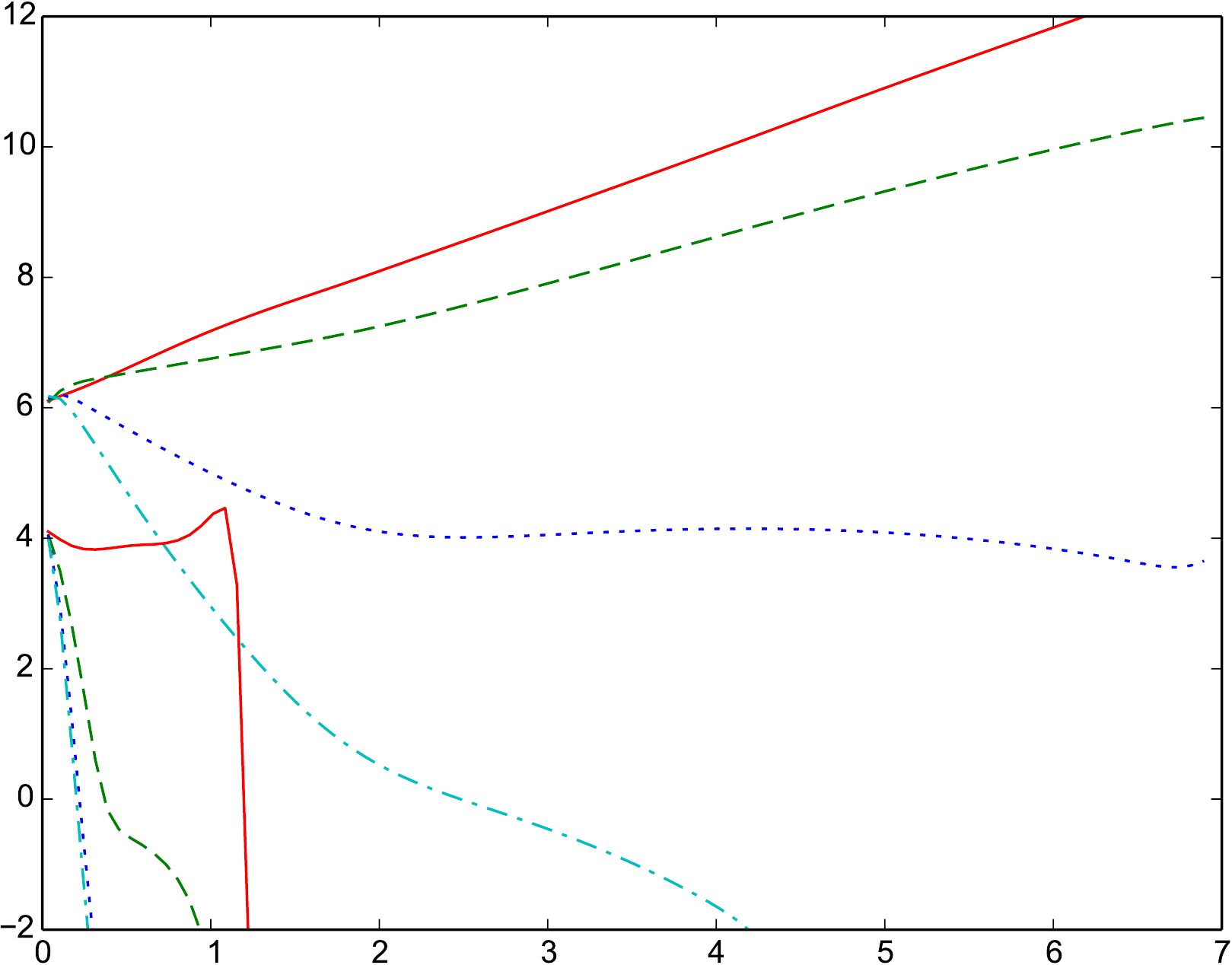}
		\put(-350,40){\rotatebox{90}{\scriptsize$\log(\;E^2j\;\mathrm{in\;MeV\; cm^{-3}})$}}
		\put(-210,-20){\scriptsize $\log(\;E\;\mathrm{in\;MeV})$}
		\put(-290,210){a)}
	\end{picture}	
		\begin{picture}(320,280)(0,0)
		\includegraphics[width=320\unitlength]{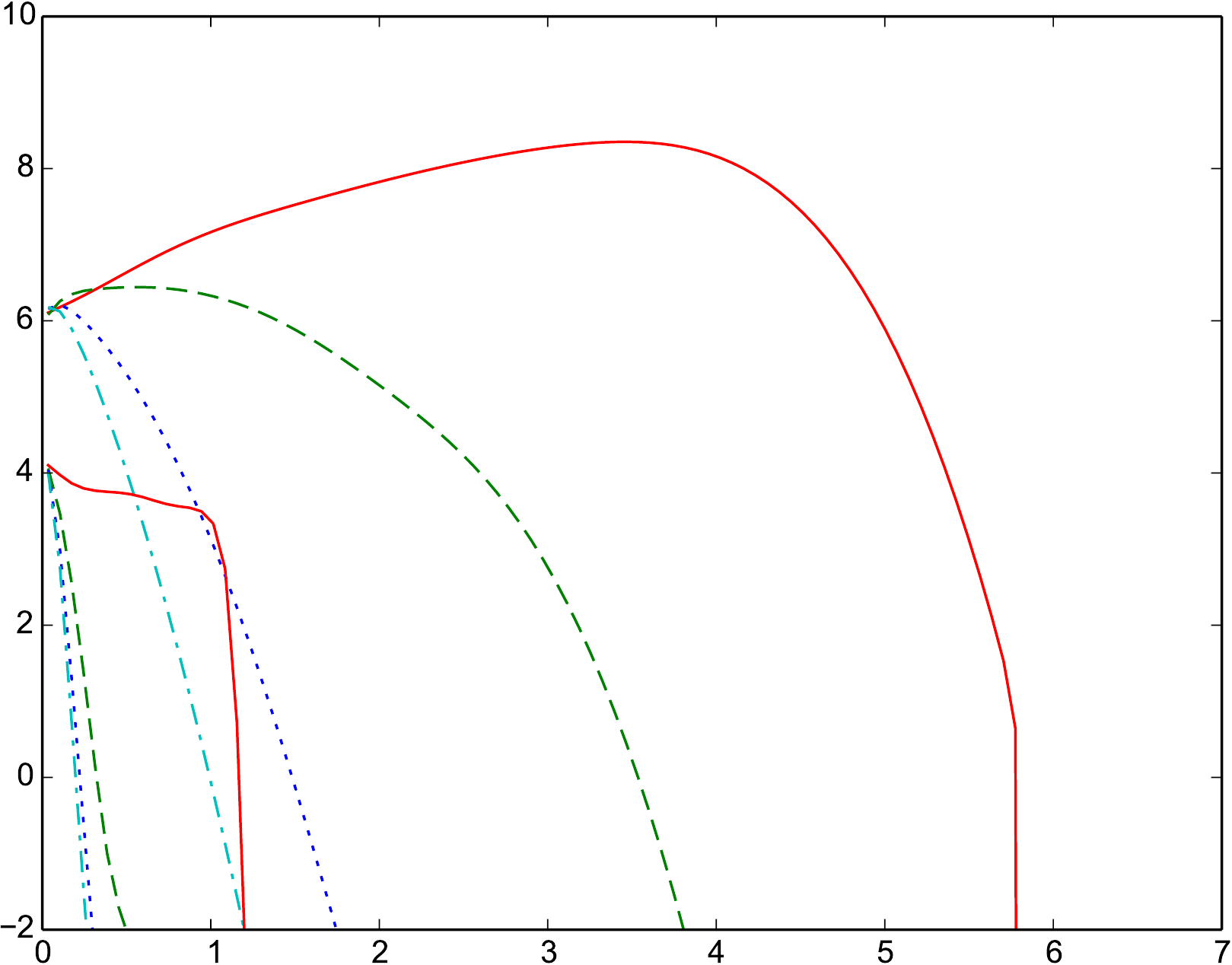}
		\put(-210,-20){\scriptsize $\log(\;E\;\mathrm{in\;MeV})$}
		\put(-290,210){b)}
	\end{picture}
		\begin{picture}(320,280)(0,0)
		\includegraphics[width=320\unitlength]{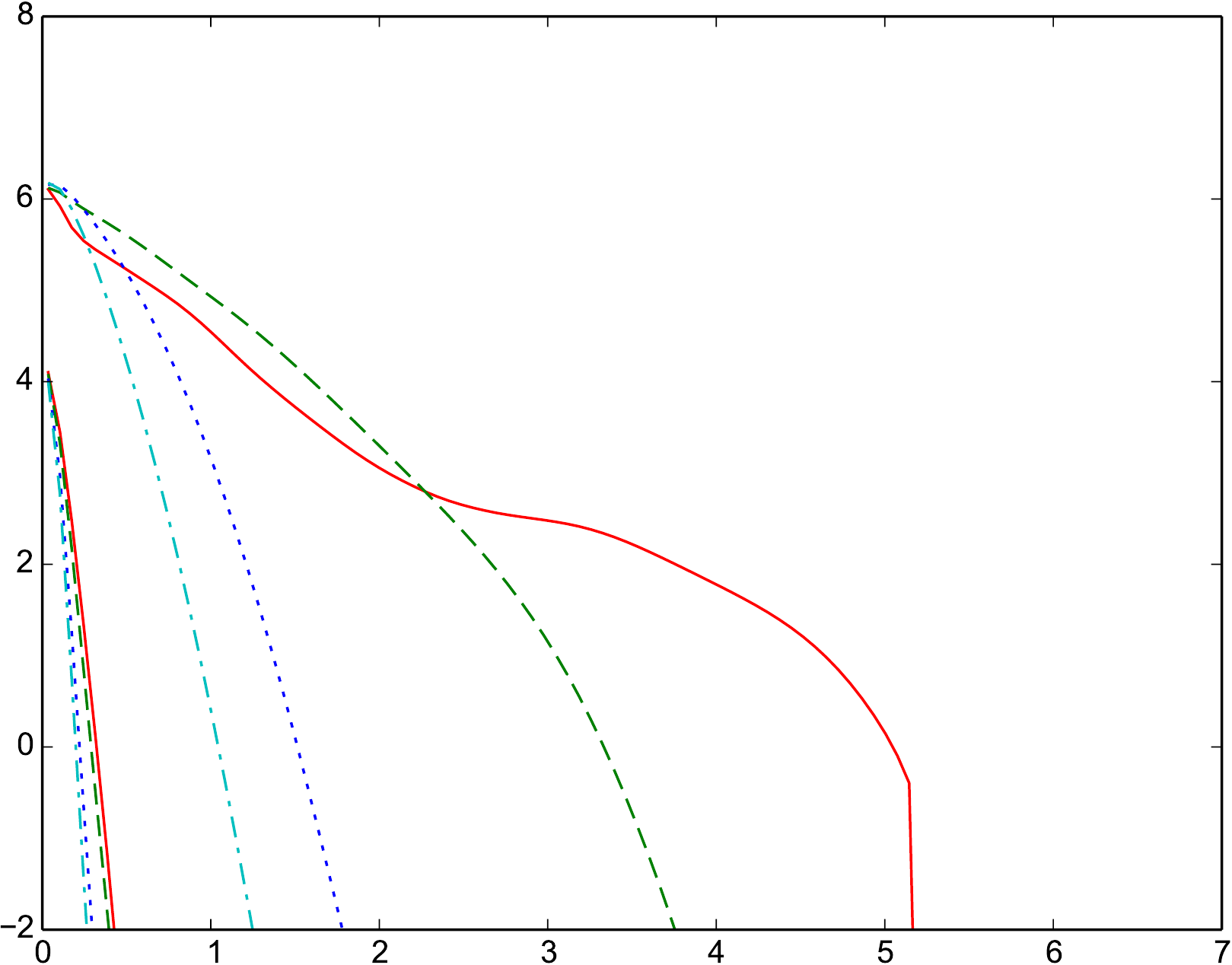}
		\put(-210,-20){\scriptsize $\log(\;E\;\mathrm{in\;MeV})$}
		\put(-290,210){c)}
	\end{picture}\\
	\begin{picture}(320,280)(0,0)
		\includegraphics[width=320\unitlength]{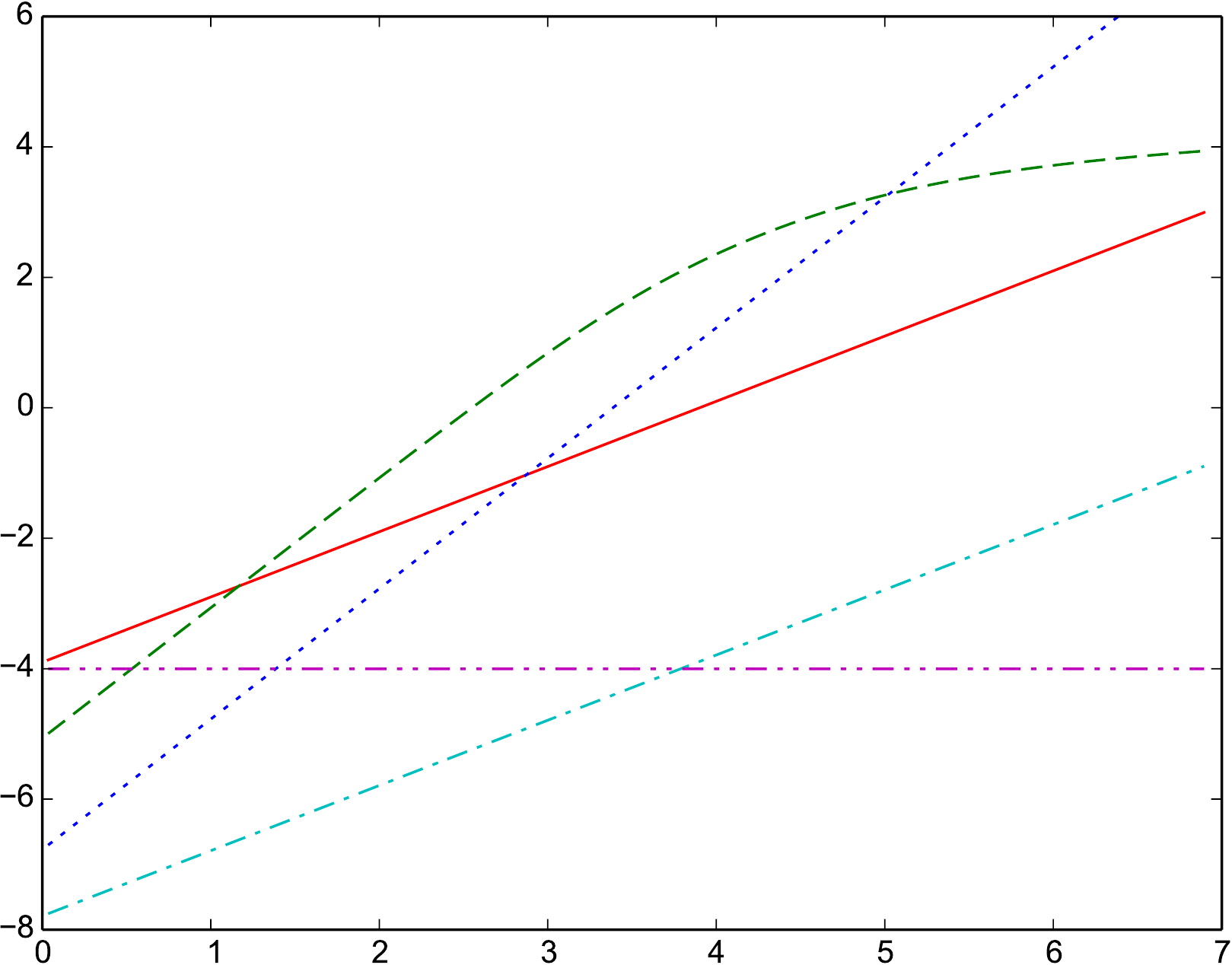}
		\put(-350,40){\rotatebox{90}{\scriptsize$\log(\;dE/dt\;\mathrm{in\;MeV\; s^{-1}})$}}
		\put(-210,-20){\scriptsize $\log(\;E\;\mathrm{in\;MeV})$}
		\put(-290,210){d)}
	\end{picture}	
		\begin{picture}(320,280)(0,0)
		\includegraphics[width=320\unitlength]{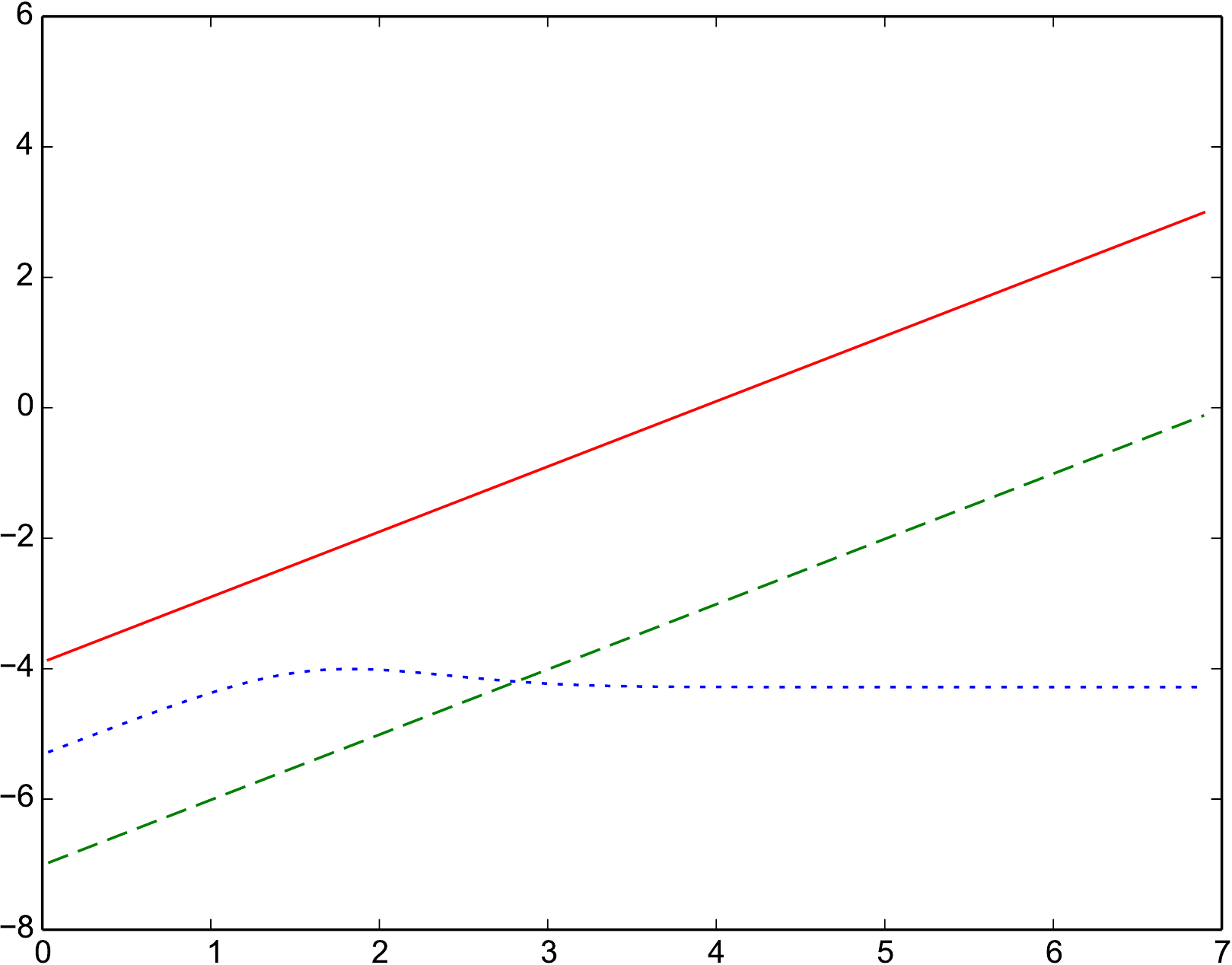}
		\put(-210,-20){\scriptsize $\log(\;E\;\mathrm{in\;MeV})$}
		\put(-290,210){e)}
	\end{picture}
\end{center}
	\caption{Upper row: Differential number density of electrons and protons for different values of the diffusion coefficient $D_0$. Electrons and protons can be distinguished by the different injection ratios (higher for protons) at $E=1\mathrm{MeV}$. The values for $D_0$ are $10^{14}$ (red solid), $10^{15}$ (green dashed), $5\times10^{15}$ (blue dotted), and $10^{16}$ m$^2$ s$^{-1}$ (azur dot--dashed). We used $\delta=0$ and no advection for plot a), $\delta=0.3$ and no advection for plot b), $\delta=0.3$ and advection for plot c).\\
	Lower row: Energy loss-rates for electrons in plot d) and for protons in plot e). The colors indicate the acceleration rate (red solid), inverse Compton losses (green dashed), synchrotron losses (blue dotted), bremsstrahlung losses (azur dot--dashed), and Coulomb losses (magenta double-dot--dashed) in plot d) and acceleration rate (red solid), losses by nucleon-nucleon collisions (green dashed), and Coulomb losses (blue dotted) in plot e).}
	\label{Diff}
\end{figure}
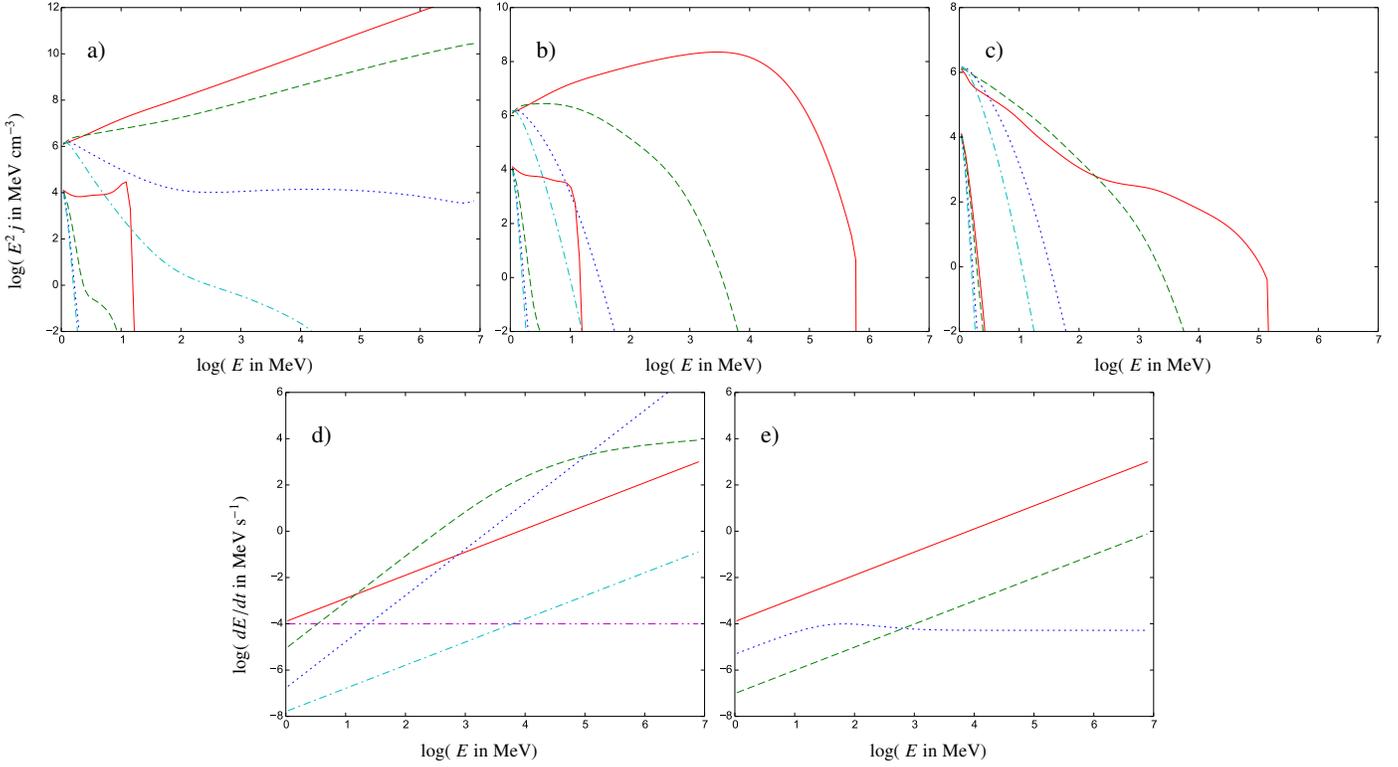

Fig. \ref{Diff} shows the resulting proton and electron spectra at the apex of the WCR in $\gamma^2$ Velorum for different values of $D_0$. For the results shown in Fig. \ref{Diff} a) diffusion is constant ($\delta=0, D=D_0$). In Fig. \ref{Diff} b) and c) the diffusion follows a power law in energy with index 0.3 ($D(E)=E^{0.3}$). Fig. \ref{Diff} c) is the only one where the advection of the particles with the flow of the wind is taken into account. It has been deactivated for Fig. \ref{Diff} a) and \ref{Diff} b) in order to make the effects of different diffusion setups visible. Below these three particle density plots, two auxiliary plots show the various energy-loss and energy-gain rates that are taken into account at the location for which the spectra are shown.\\

The proton spectra of Fig.~\ref{Diff}~a) indicate that -- without the implementation of either Bohm-diffusion or energy-dependent diffusion -- the spectra have no cutoff. This is also apparent from the relative strengths of the energy-loss and gain-rates in Fig.~\ref{Diff}~e). The acceleration (solid red) dominates the combined losses for all energies. The spectral index can be controlled by the normalization of the diffusion coefficient $D_0$. In our model we find that for $D_0\geq 10^{16}$ m$^2$ s$^{-1}$ and $\delta=0.3$, protons will not reach energies above 10~GeV. At this energy they simply diffuse out of the region in which further acceleration were possible. The change in spectral index at $\sim$100 MeV -- most visible in the two spectra for stronger diffusion in Fig.~\ref{Diff}~a) -- can be readily explained by the change in Coulomb losses for protons which roughly become constant above this energy. As the difference between energy gains and losses in Fig. \ref{Diff}~e) increases, the proton spectra become harder.\\

For the electrons in Fig. \ref{Diff} a) the situation is very different. The cutoff reached just above 10 MeV for the case of low diffusion is caused by the strong inverse Compton losses in the vicinity of the O~star. Plot d) illustrates that acceleration is dominant only in a small energy range below 10 MeV. For the cases of higher diffusion we find - as for protons - a softening of the spectra. Together with the severe energy losses this leads to an earlier cutoff.\\

Note that the situation for the electrons changes somewhat as larger distances from the stars are reached in the outer wings of the WCR. As the effect of inverse Compton losses subsides, electron energies of several 100 MeV can be reached. However - considering the magnetic dipole field - synchrotron losses will become dominant in the direction of the poles of the stellar dipoles and lead to an even lower energetic cutoff than caused by inverse Compton losses. While an accurate representation of the magnetic field (as provided by the full-MHD simulation) is not of vital importance near the apex, where losses linked to the stellar radiation fields dominate, it is of high relevance at other regions of the WCR. \\

In Fig. \ref{Diff} b) the diffusion is not constant anymore but grows with $E^{0.3}$. This produces a cutoff -- even if the diffusion is low at low energies. The graph illustrates that an implementation of a Bohm-diffusion related cutoff is not needed anymore for the case of energy-dependent diffusion. It also emphasizes the role of $D_0$ and $\delta$ as important parameters to determine the maximum energy reached by protons. For electrons these parameters would only become relevant in case of very low loss-rates as might be the case for systems where the WCR is far from the stars, but certainly not in $\gamma^2$~Velorum.\\

Fig. \ref{Diff} c) includes the effect of advection whereas in the previous discussion, the particles have been allowed to diffuse freely away from the shock front without being affected by the flow of the wind. It has to be stressed that the WCR in a CWB system cannot be described by the properties of a typical 1D shock. An important difference is the significant velocity component perpendicular to the shock normal in the post-shock wind which leads to particle transport in downstream direction. The effects are seen in the figure. Including advection we find a significant softening of all spectra. However - the maximum energies that are reached are still very close to the case without advection in it. Although the spectra are much softer at low energies (compared to Fig. \ref{Diff} b)), the spectral indices are similar at higher energies where diffusion dominates and the effects of advection disappear. The apparent irregularities for the case of very low diffusion (wiggles in the red solid spectra in Fig. \ref{Diff} c) ) are explained by the local turbulence of the plasma which leaves an imprint on the spectra. This effect can be seen for low diffusion and is enhanced if advection is dominant. For the spectra with $D_0=10^{14}$ m$^2$s$^{-1}$ in Fig. \ref{Diff} c), the effect of the vanishing influence of Compton losses and the corresponding hardening of the spectrum above $\sim$100 MeV becomes apparent once again. \\

\begin{figure}
	\setlength{\unitlength}{0.0008\textwidth}
		\begin{picture}(335,321)(0,0)
		\put(0,0){\includegraphics[height=320\unitlength,trim=0cm 0cm 2.2cm 0 cm, clip=true]{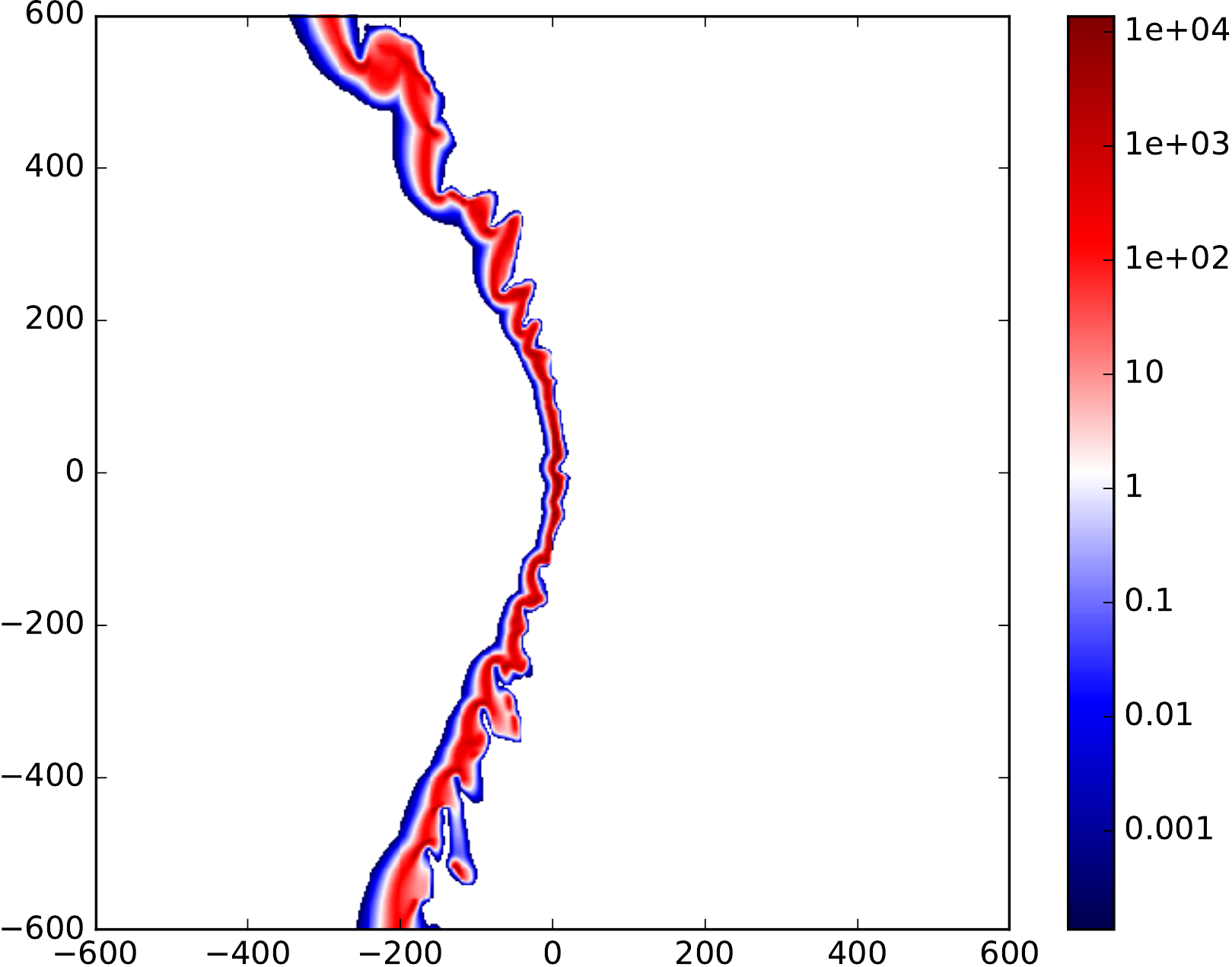}}
		\put(60,130){\large 1 MeV}
		\put(0,150){\rotatebox{90}{\scriptsize $z\;[\mathrm{R}_\odot] $}}
	\end{picture}	
		\begin{picture}(50,321)(0,0)
		\put(0,0){\includegraphics[height=320\unitlength,trim=14cm 0cm 0cm 0 cm, clip=true]{./ap_el1_xz.pdf}}
	\end{picture}
		\begin{picture}(335,321)(0,0)
		\put(0,0){\includegraphics[height=320\unitlength,trim=0cm 0cm 2.2cm 0 cm, clip=true]{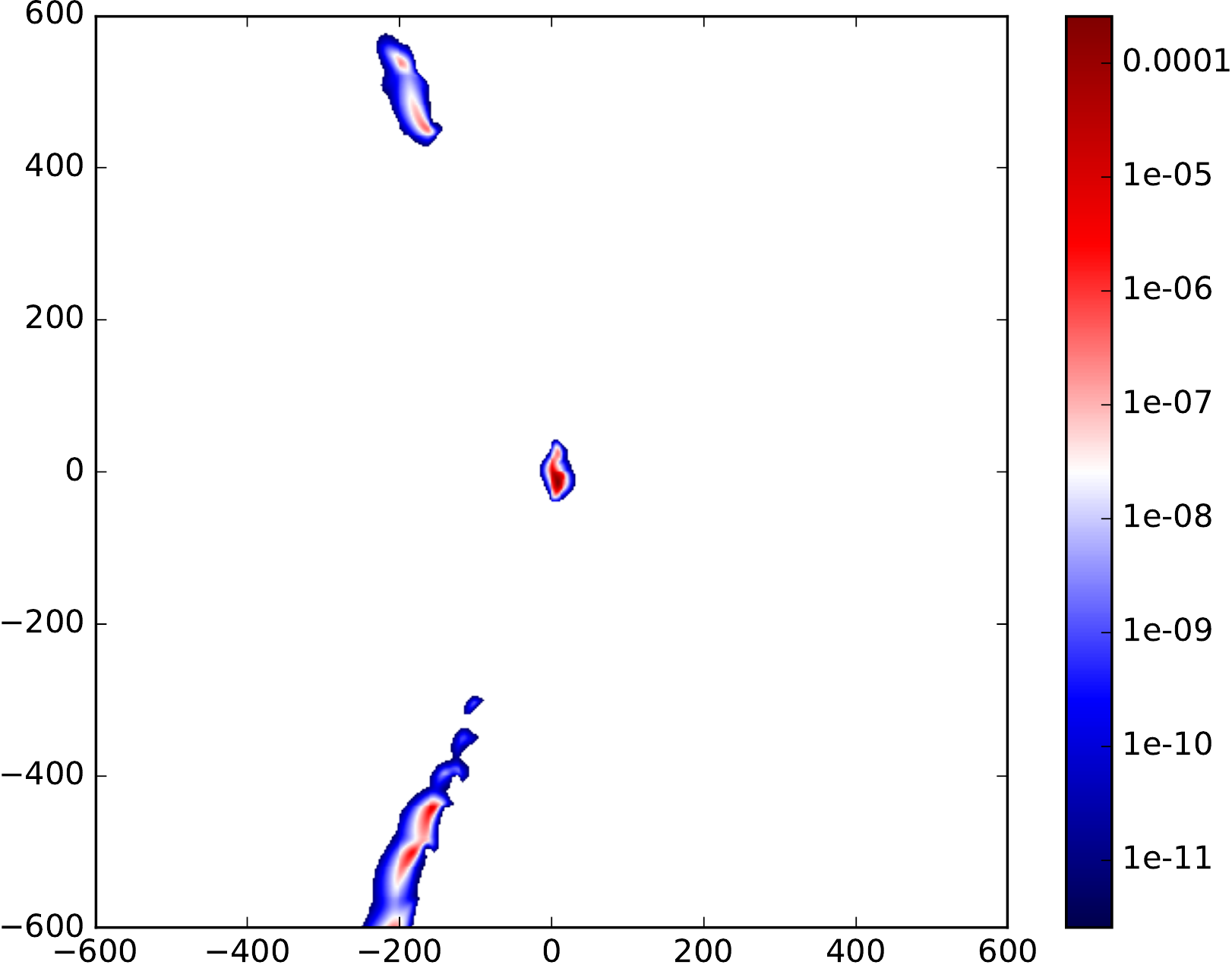}}
		\put(60,130){\large 10 MeV}
		\put(0,150){\rotatebox{90}{\scriptsize $z\;[\mathrm{R}_\odot] $}}
	\end{picture}		
		\begin{picture}(50,321)(0,0)
		\put(0,0){\includegraphics[height=320\unitlength,trim=14cm 0cm 0cm 0 cm, clip=true]{./ap_el10_xz.pdf}}
	\end{picture}
		\begin{picture}(335,321)(0,0)
		\put(0,0){\includegraphics[height=320\unitlength,trim=0cm 0cm 2.2cm 0 cm, clip=true]{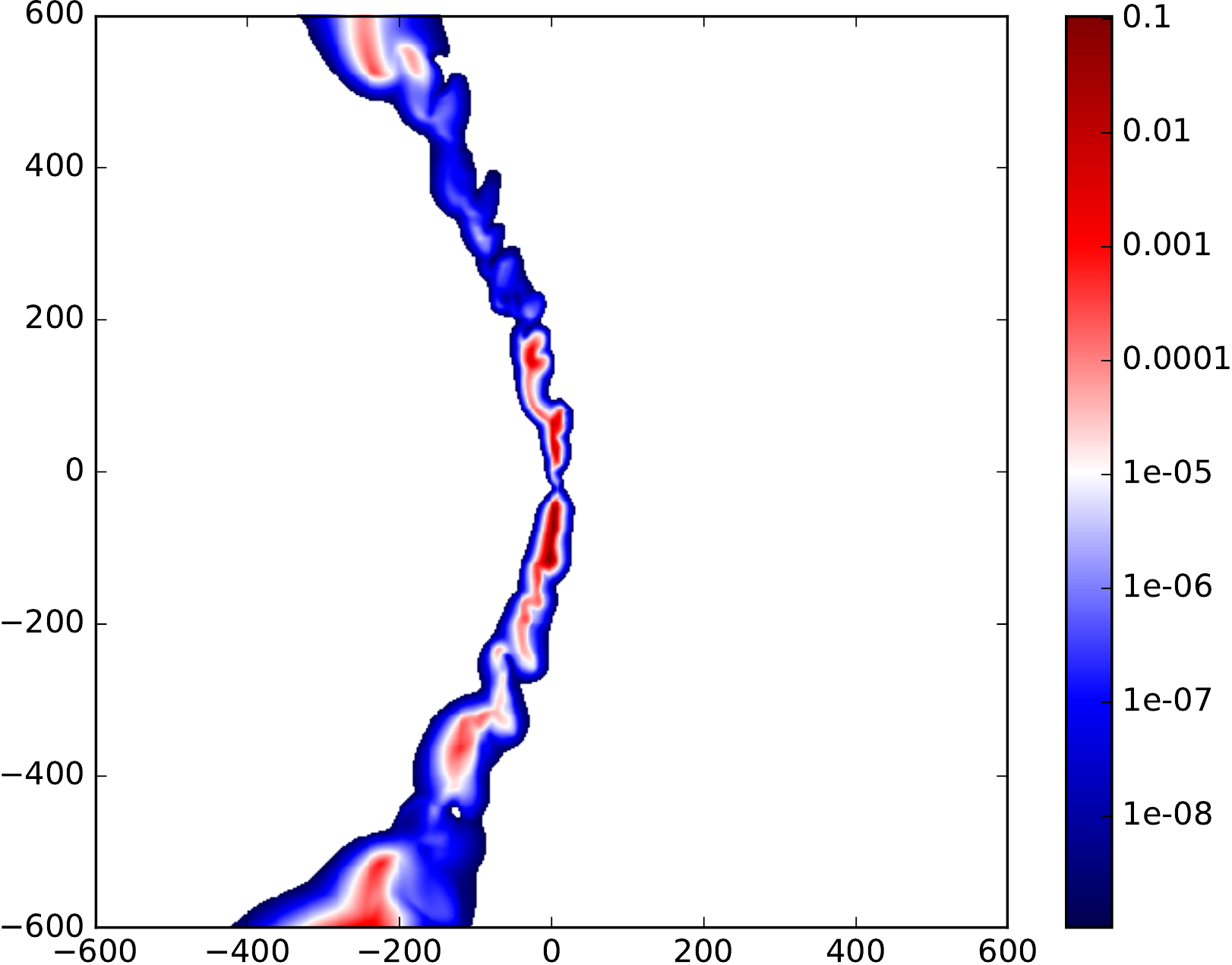}}
		\put(60,130){\large 10 MeV}
		\put(0,150){\rotatebox{90}{\scriptsize $y\;[\mathrm{R}_\odot] $}}
	\end{picture}			
		\begin{picture}(50,321)(0,0)
		\put(0,0){\includegraphics[height=320\unitlength,trim=14cm 0cm 0cm 0 cm, clip=true]{./ap_el10_xy.pdf}}
	\end{picture}\\
			\begin{picture}(335,321)(0,0)
		\put(0,0){\includegraphics[height=320\unitlength,trim=0cm 0cm 2.2cm 0 cm, clip=true]{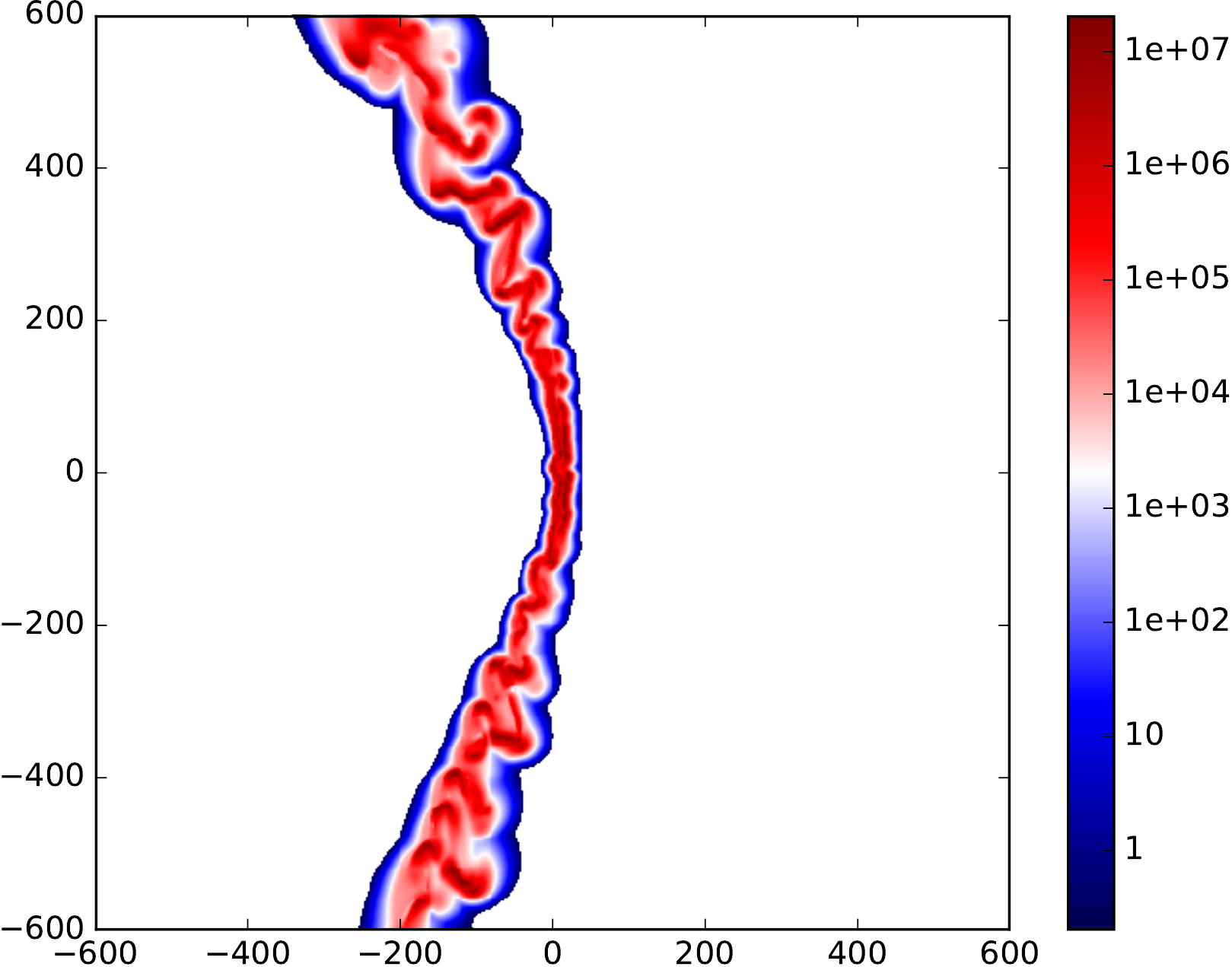}}
		\put(60,130){\large 1 MeV}
		\put(0,150){\rotatebox{90}{\scriptsize $z\;[\mathrm{R}_\odot] $}}
		\put(150,-15){\scriptsize $x\;[\mathrm{R}_\odot] $}
	\end{picture}	
		\begin{picture}(50,321)(0,0)
		\put(0,0){\includegraphics[height=320\unitlength,trim=14cm 0cm 0cm 0 cm, clip=true]{./ap_pr1_xz.pdf}}
		\put(-5,-15){\scriptsize MeV$^{-1}$ m$^{-3}$}
	\end{picture}
		\begin{picture}(335,321)(0,0)
		\put(0,0){\includegraphics[height=320\unitlength,trim=0cm 0cm 2.2cm 0 cm, clip=true]{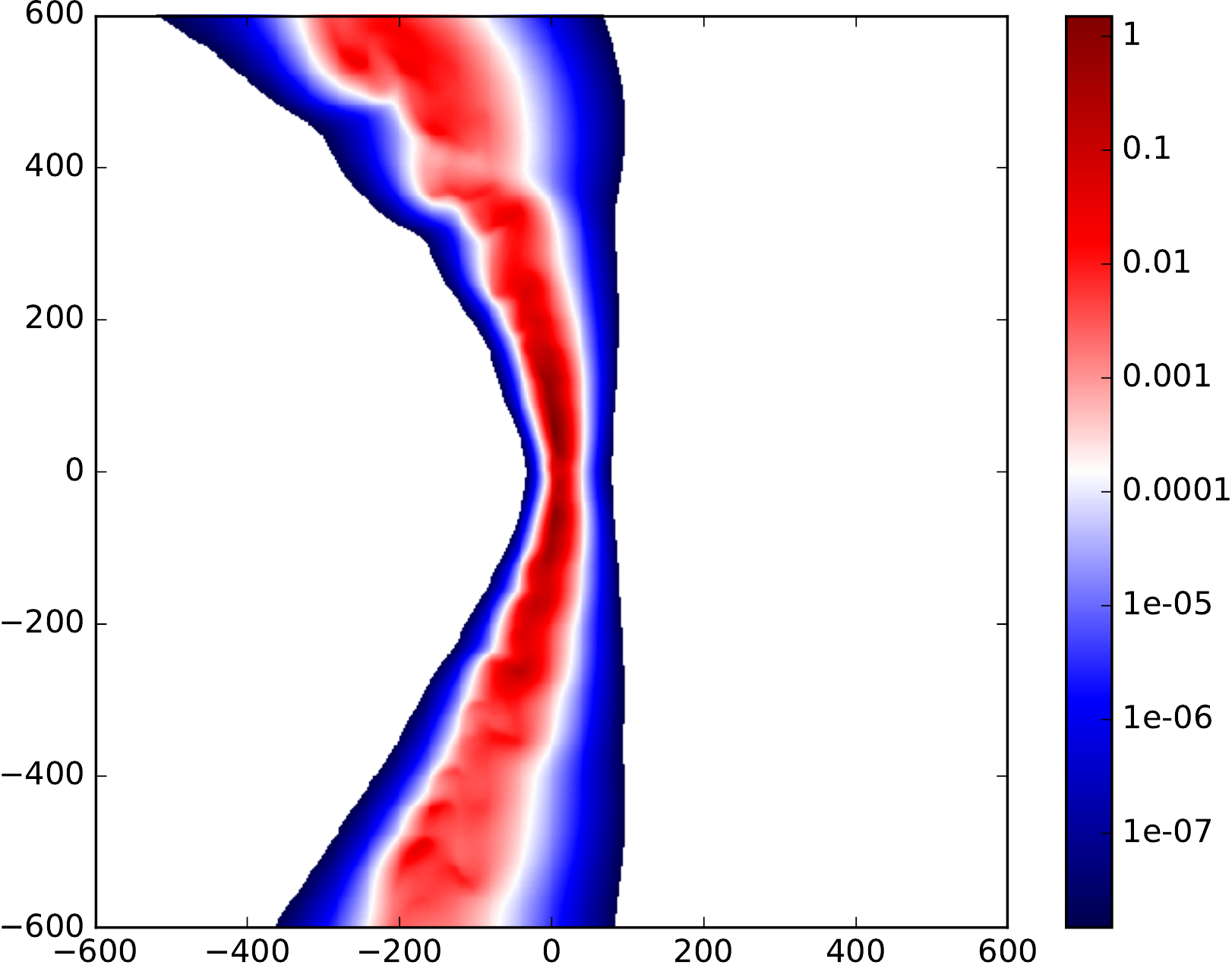}}
		\put(60,130){\large 100 MeV}
		\put(150,-15){\scriptsize $x\;[\mathrm{R}_\odot] $}
	\end{picture}		
		\begin{picture}(50,321)(0,0)
		\put(0,0){\includegraphics[height=320\unitlength,trim=14cm 0cm 0cm 0 cm, clip=true]{./ap_pr100_xz.pdf}}
		\put(-5,-15){\scriptsize MeV$^{-1}$ m$^{-3}$}
	\end{picture}
		\begin{picture}(335,321)(0,0)
		\put(0,0){\includegraphics[height=320\unitlength,trim=0cm 0cm 2.2cm 0 cm, clip=true]{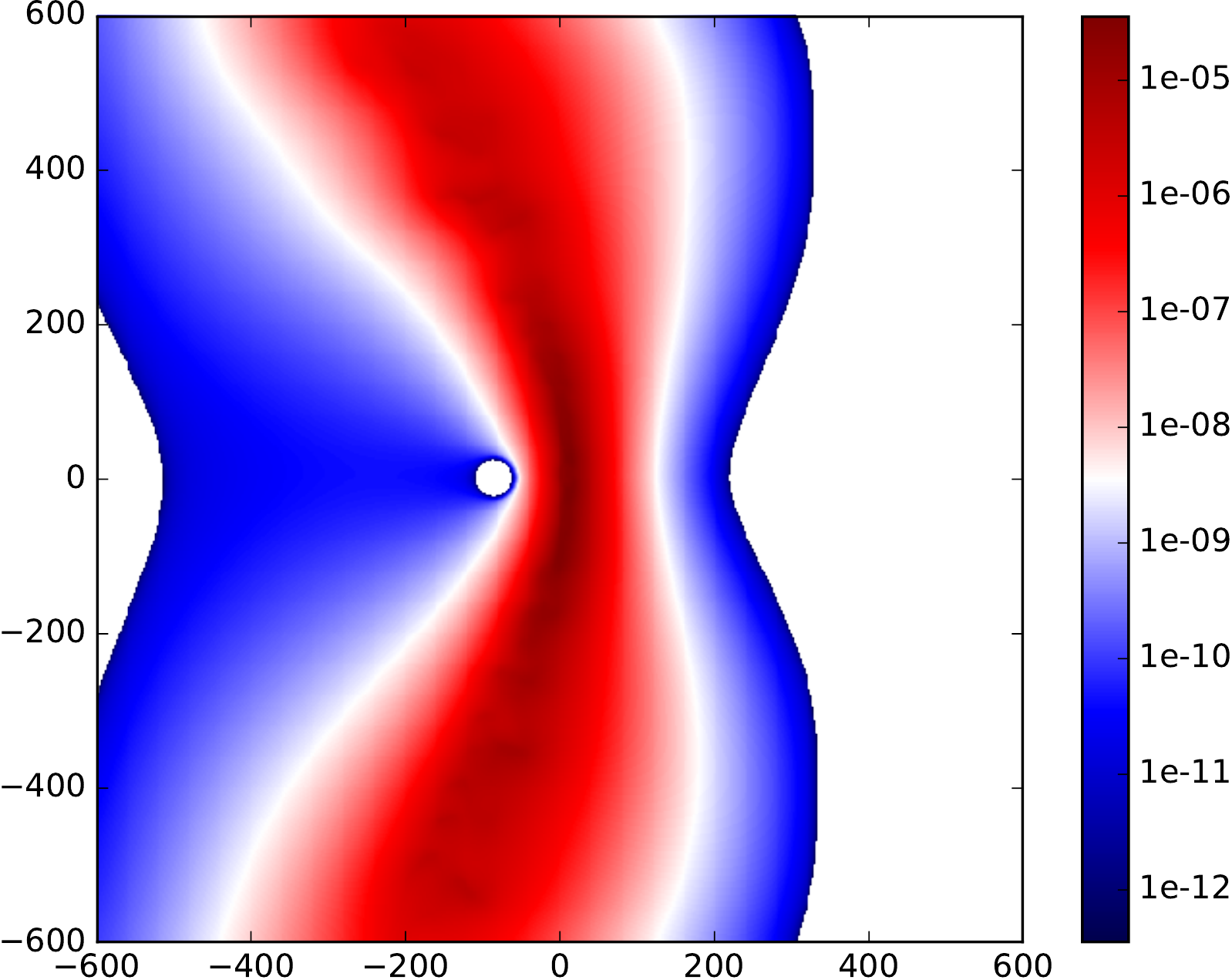}}
		\put(60,130){\large 10 GeV}
		\put(150,-15){\scriptsize $x\;[\mathrm{R}_\odot] $}
	\end{picture}			
		\begin{picture}(50,321)(0,0)
		\put(0,0){\includegraphics[height=320\unitlength,trim=14cm 0cm 0cm 0 cm, clip=true]{./ap_pr10000_xz.pdf}}
		\put(-5,-15){\scriptsize MeV$^{-1}$ m$^{-3}$}
	\end{picture}
	
	\caption{Differential number density of electrons (top) and protons (bottom) in MeV$^{-1}$ m$^{-3}$ for different values of kinetic particle energy. The countor maps show the x--z plane of a 512$^3$ simulation at y=0.}
	\label{partexp}
\end{figure}

Fig. \ref{partexp} explores the spatial distribution of electrons and protons at different energies. To simplify matters only apastron conditions are shown here.
Apparently, the strong stellar radiation field close to the apex of the WCR leads to severe inverse Compton losses which prevent electrons from reaching higher energies than $\sim$10 MeV at the apex and $\sim$100 MeV in the outer wings. 
The proton densities illustrate the effect of energy dependent diffusion. The higher the particle energy gets, the larger the populated region becomes. The turbulent structure of the apastron WCR disappears at higher energies as small-scale variations are smoothed out by the dominant diffusion. \\

\begin{figure}
	\setlength{\unitlength}{0.0008\textwidth}
		\begin{picture}(335,321)(0,0)
		\put(0,0){\includegraphics[height=320\unitlength,trim=0cm 0cm 2.2cm 0 cm, clip=true]{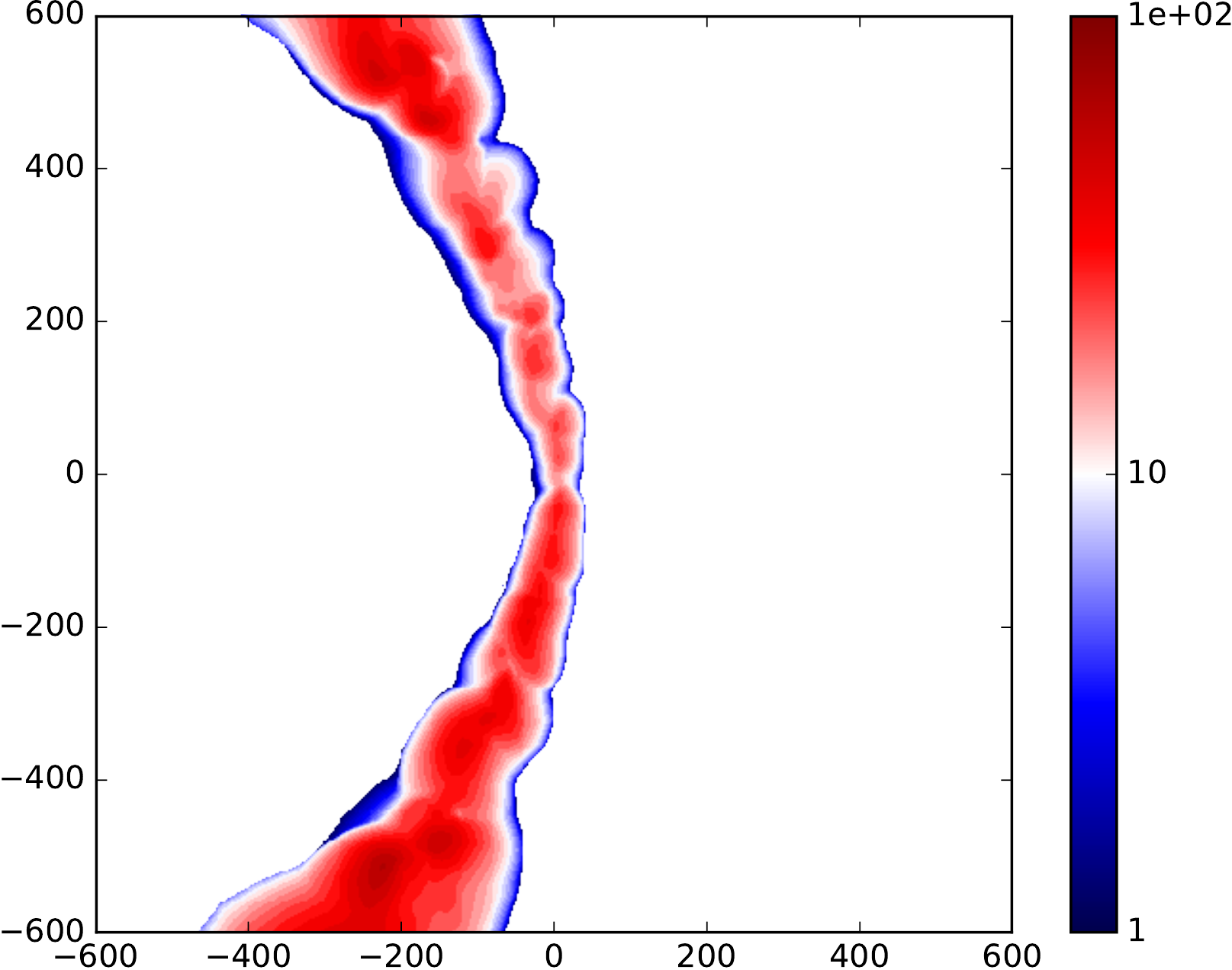}}
		\put(0,150){\rotatebox{90}{\scriptsize $y\;[\mathrm{R}_\odot] $}}
		\put(150,-15){\scriptsize $x\;[\mathrm{R}_\odot] $}
	\end{picture}	
		\begin{picture}(50,321)(0,0)
		\includegraphics[height=320\unitlength,trim=14cm 0cm 0cm 0 cm, clip=true]{./ap_logemax.pdf}
		\put(-60,-15){\scriptsize $E_\mathrm{max}\;[\mathrm{MeV}] $}
	\end{picture}
		\begin{picture}(335,321)(0,0)
		\put(0,0){\includegraphics[height=320\unitlength,trim=0cm 0cm 2.2cm 0 cm, clip=true]{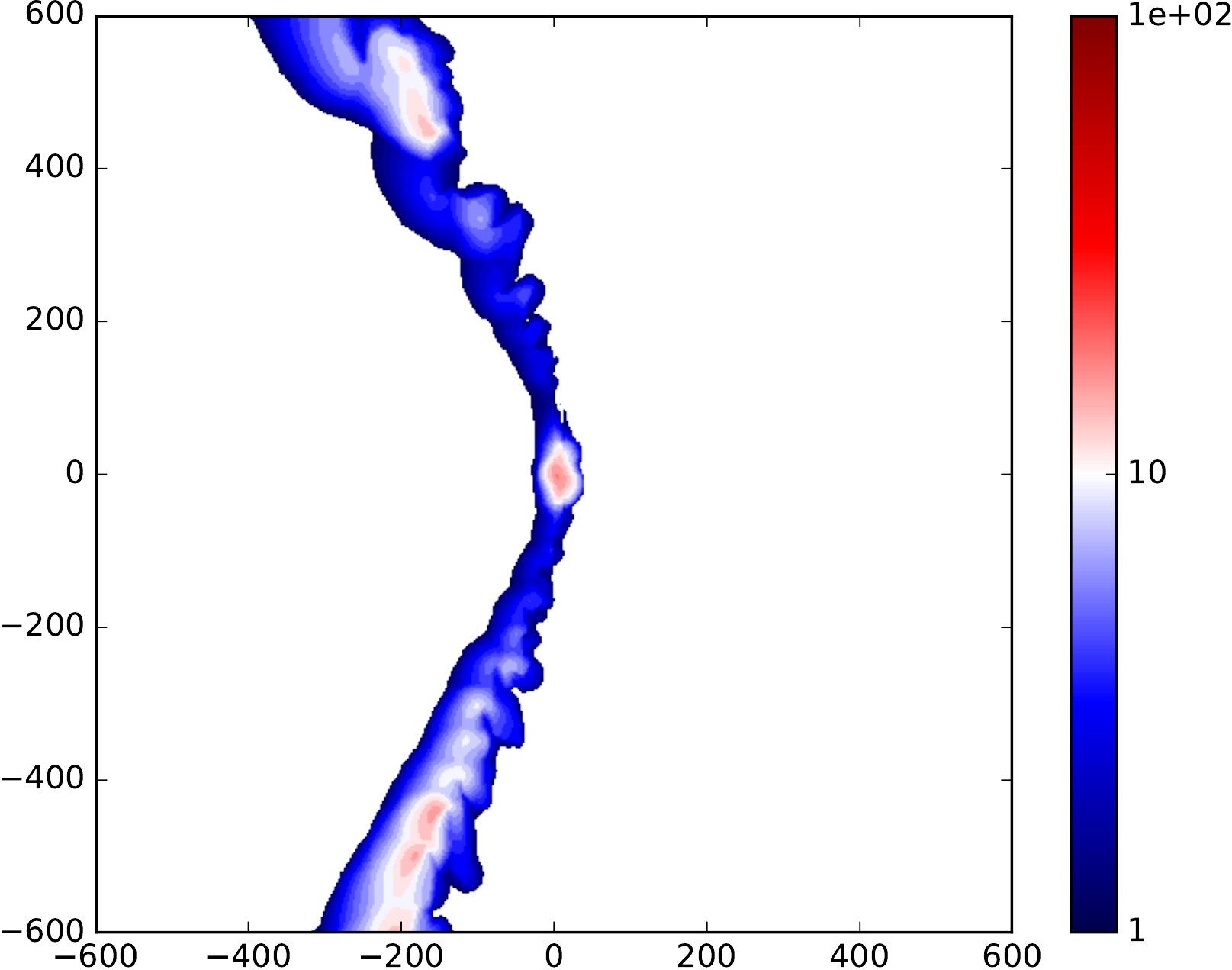}}
		\put(0,150){\rotatebox{90}{\scriptsize $z\;[\mathrm{R}_\odot] $}}
		\put(150,-15){\scriptsize $x\;[\mathrm{R}_\odot] $}
	\end{picture}		
		\begin{picture}(50,321)(0,0)
		\includegraphics[height=320\unitlength,trim=14cm 0cm 0cm 0 cm, clip=true]{./ap_logemax_xz.pdf}
		\put(-60,-15){\scriptsize $E_\mathrm{max}\;[\mathrm{MeV}] $}
	\end{picture}
		\begin{picture}(335,321)(0,0)
		\put(0,0){\includegraphics[height=320\unitlength,trim=0cm 0cm 2.2cm 0 cm, clip=true]{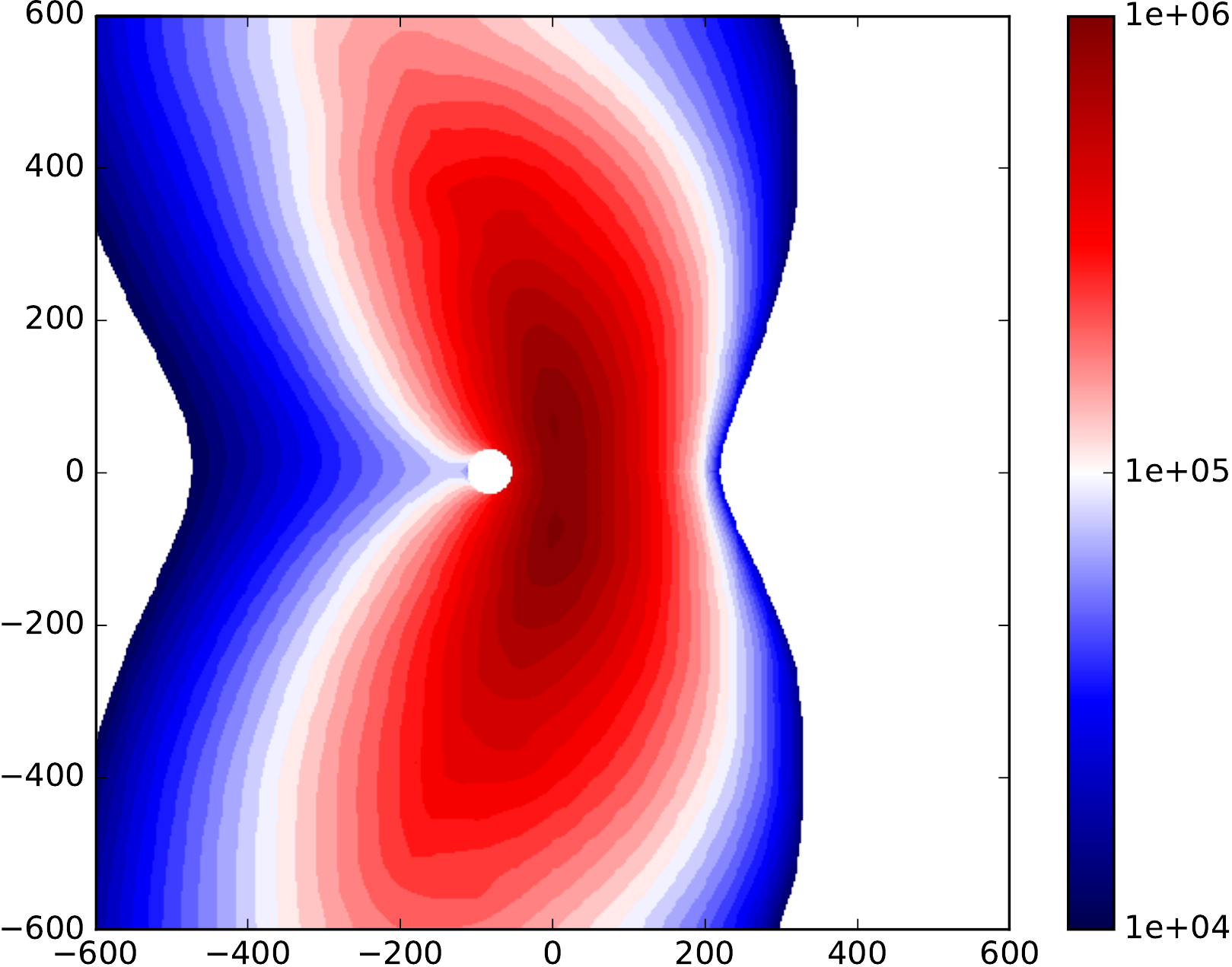}}
		\put(0,150){\rotatebox{90}{\scriptsize $z\;[\mathrm{R}_\odot] $}}
		\put(150,-15){\scriptsize $x\;[\mathrm{R}_\odot] $}
	\end{picture}			
		\begin{picture}(50,321)(0,0)
		\includegraphics[height=320\unitlength,trim=14cm 0cm 0cm 0 cm, clip=true]{./ap_logpmax.pdf}
		\put(-66,-15){\scriptsize $E_\mathrm{max}\;[\mathrm{MeV}] $}
	\end{picture}
	\caption{Maximum particle energies for electrons in the $y-z$ plane (left), electrons in the x--z plane (middle) and protons in the x--z plane (right).}
	\label{maxes}
\end{figure}

Fig. \ref{maxes} shows the highest particle  energies reached. A clear difference can be seen between the electron energies in the x--y (left) and x--z (centre) plane. This is due to the effect of synchrotron losses caused by the dipolar shape of the magnetic field. Around the dipole equator in the x-y plane field strengths are low and electrons therefore reach higher energies. In the center plot, large parts of the WCR are in close vicinity to the lobes of the dipole field which cause severe losses by synchrotron emission. Electron energies decrease accordingly. For protons there is no such effect. Diffusion allows for a population of protons up to $\sim$1~TeV around the apex of the WCR where the acceleration of particles is most efficient. In regions further downstream, the maximum energies are generally lower. This is a combined effect of less acceleration due to lower velocity gradients at the shock and of energy loss by collisions and Coulomb losses as the protons move downstream. \\

Fig. \ref{SEDs} (left) shows the resulting particle spectra when integrating over the simulated volume. The agreement with the previously discussed single-cell spectra and contour maps of particle densities is apparent. 

\subsection{Gamma-ray predictions}
\begin{figure}
	\setlength{\unitlength}{0.0011\textwidth}
\begin{picture}(450,350)(0,0)
		\put(0,30){\includegraphics[height=280\unitlength,trim=0cm 0cm 0cm 0 cm, clip=true]{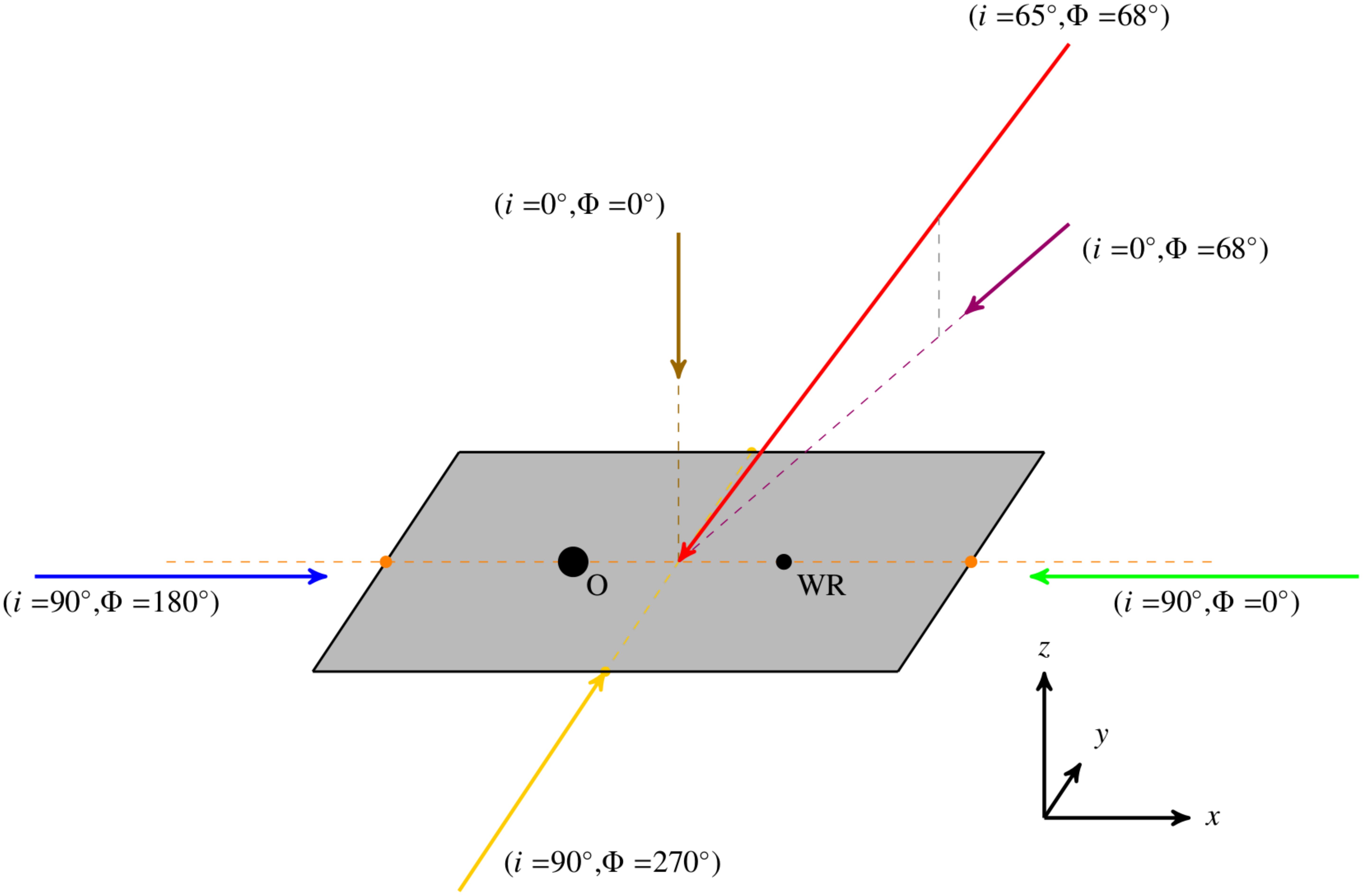}}
	\end{picture}
	\begin{picture}(335,321)(0,0)
		\put(0,0){\includegraphics[height=320\unitlength,trim=1.2cm 0.4cm 2.4cm 0 cm, clip=true]{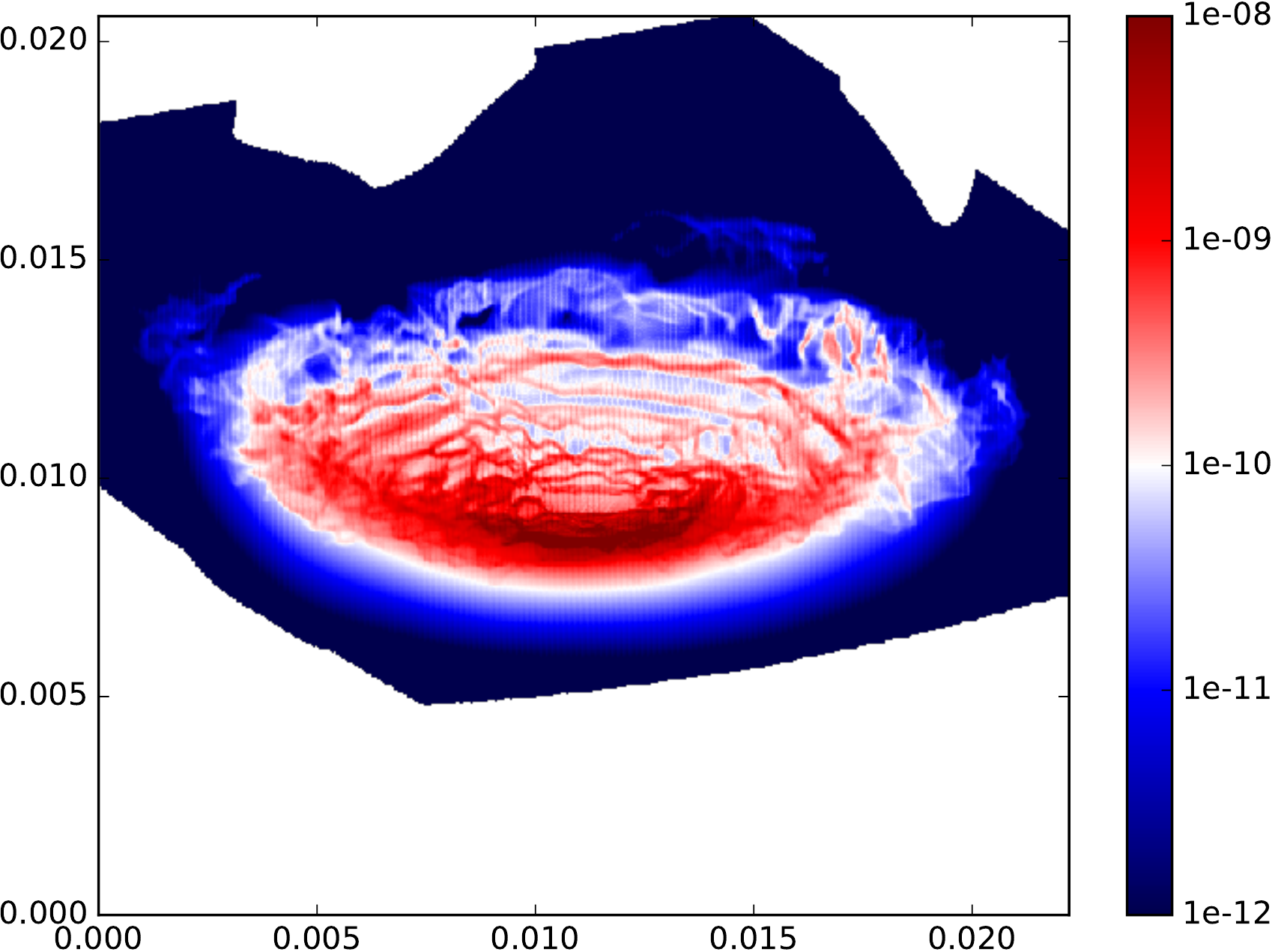}}
		\put(135,20){\line(1,0){79}}
		\put(135,17){\line(0,1){6}}
		\put(214,17){\line(0,1){6}}
		\put(135,25){\scriptsize{0.005 arcsec}}
	\end{picture}	
		\begin{picture}(0,321)(0,0)
		\put(0,15){\includegraphics[height=305\unitlength,trim=14.9cm 0.36cm 0cm 0 cm, clip=true]{p0_2.pdf}}
		\put(0,0){\scriptsize Flux [ m$^{-2}$ s$^{-1}$]}
	\end{picture}
	
			\caption{Left: 
			 view of the two stars within the computational domain. The line of centers is represented by the horizontal dashed orange line. Various viewing angles (lines-of-sight) are indicated, including the one that is used in our simulation: $i=65^\circ, \Phi=68^\circ$. Right: Projected flux above 100 MeV for neutral pion decay. }
	\label{pion}
\end{figure}

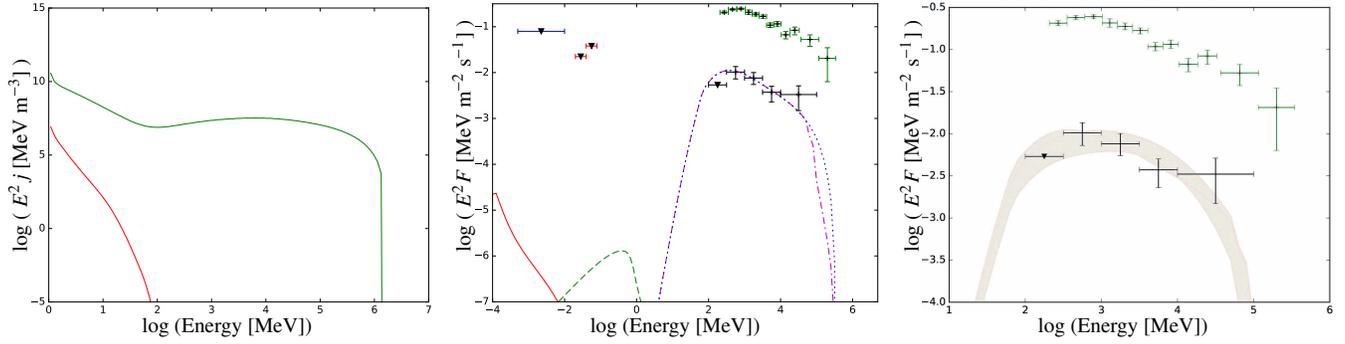
\begin{figure}
	\setlength{\unitlength}{0.00065\textwidth}
	\begin{picture}(490,350)(0,0)
		\put(25,10){\includegraphics[height=350\unitlength,trim=0cm 0.0cm 0cm 0 cm, clip=true]{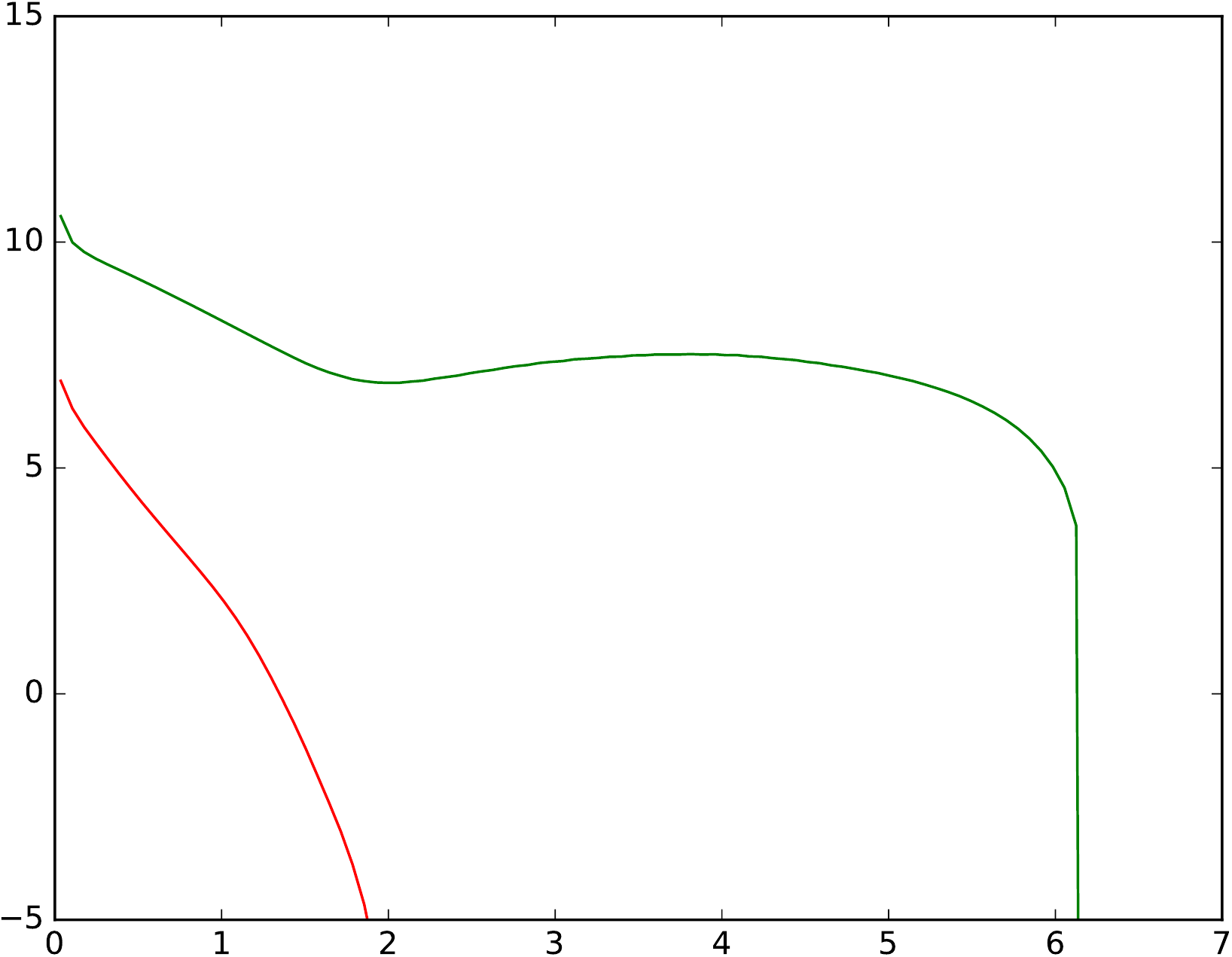}}
		\put(0,80){\rotatebox{90}{\footnotesize $\log$ ( $E^2j$ [MeV m$^{-3}$] ) }}
		\put(150,-10){\footnotesize $\log$ (Energy [MeV])}					
	\end{picture}	
		\begin{picture}(490,350)(0,0)
		\put(20,10){\includegraphics[height=350\unitlength,trim=0cm 0.0cm 0cm 0 cm, clip=true]{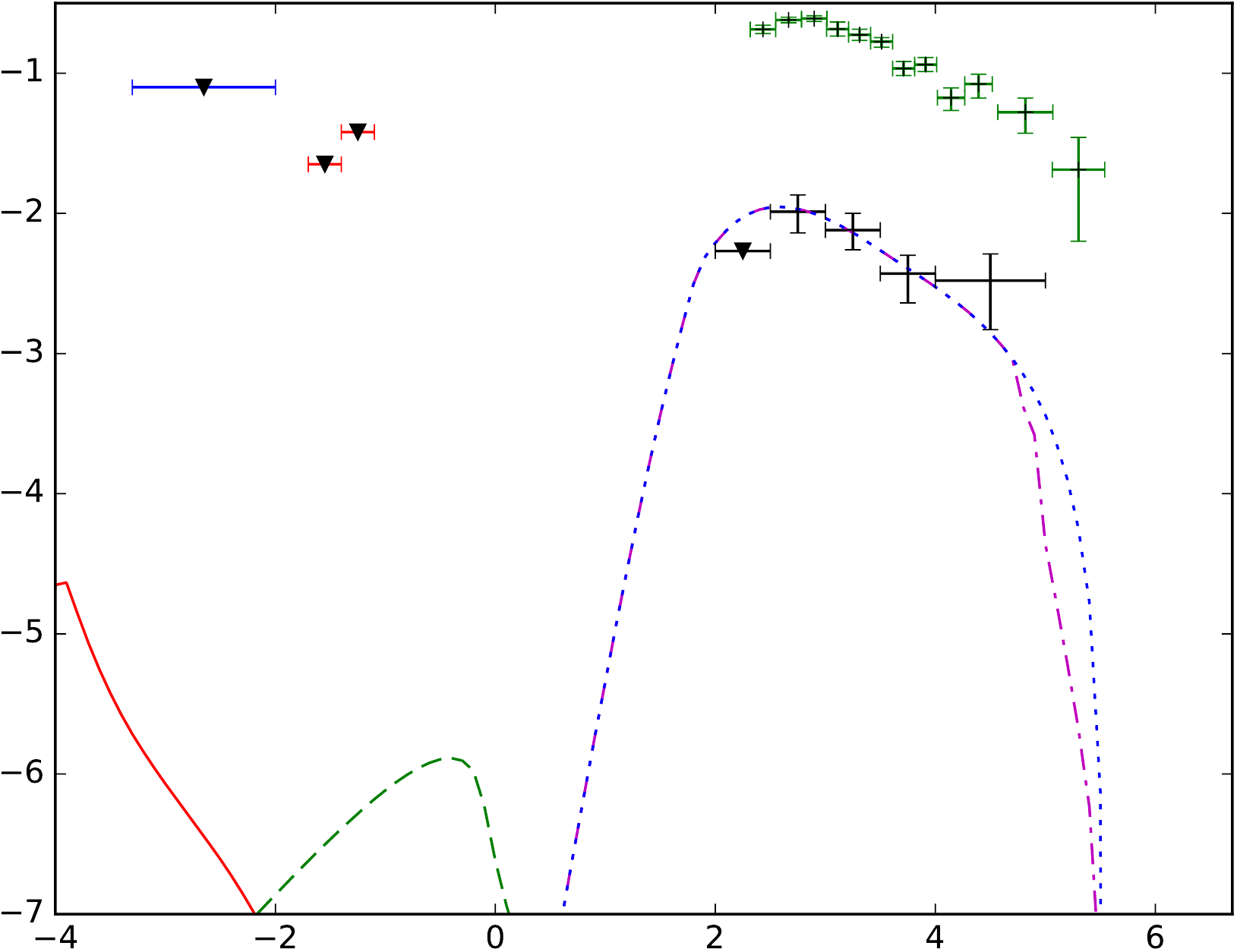}}
		\put(-5,80){\rotatebox{90}{\footnotesize $\log$ ( $E^2F$ [MeV m$^{-2}$ s$^{-1}$] ) }}
		\put(150,-10){\footnotesize $\log$ (Energy [MeV])}		
	\end{picture}
	\begin{picture}(490,350)(0,0)
		\put(20,10){\includegraphics[height=350\unitlength,trim=0cm 0.0cm 0cm 0 cm, clip=true]{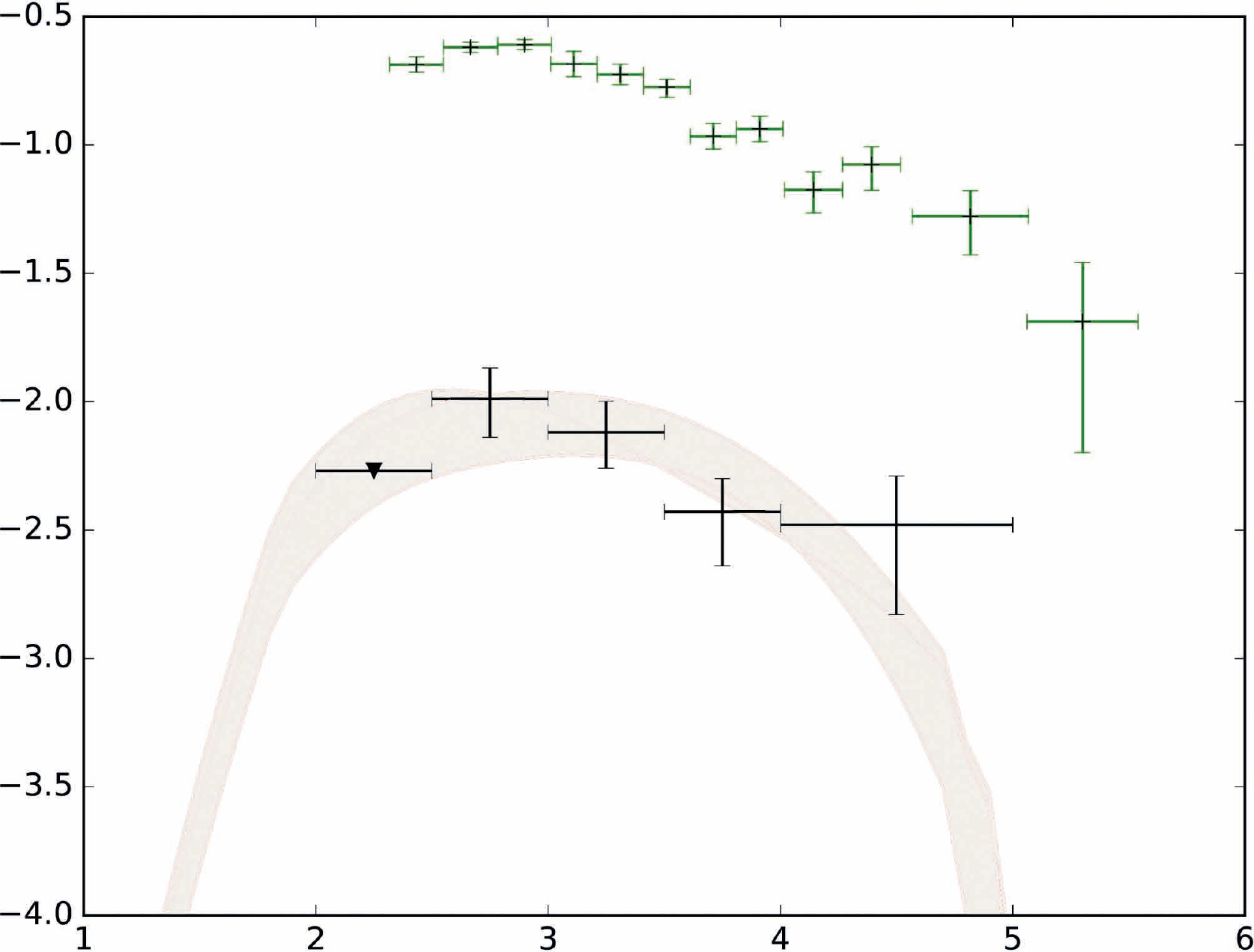}}
		\put(-5,80){\rotatebox{90}{\footnotesize $\log$ ( $E^2F$ [MeV m$^{-2}$ s$^{-1}$] ) }}
		\put(150,-10){\footnotesize $\log$ (Energy [MeV])}		
	\end{picture}
		\caption{Left: number density of protons (green) and electrons (red) integrated over the computational domain. Middle: resulting SEDs for inverse Compton (solid red), bremsstrahlung (dashed green) and neutral pion decay emission channels, the latter without (dotted blue) and with photon-photon absorption (dot-dashed magenta). Also shown are the \textit{Fermi}-LAT data points determined for $\gamma^2$ Velorum by \cite{Pshirkov2016} (black) and for $\eta$ Carinae by  \citep{Reitberger2015} (green) as well as the upper limit on the 0.5-10 keV emission of $\gamma^2$~Velorum derived from ASCA, and the INTEGRAL/IBIS upper limits \citep[for both, see ][]{Tatischeff2004}.    Right: The gray-shaded area indicates the simulated fluctuation of the neutral pion decay induced SEDs due to the turbulence in the WCR (stellar separation fixed at apastron). The data points are the same as in the center plot.}
	\label{SEDs}
\end{figure}

Processing the obtained particle spectra (for $D_0=8\times 10^{-14}$ m$^2 $s$^{-1}$, $\delta=0.3$ and proton injection ratio $\eta_p=10^{-3}$) and the MHD variables with the schemes presented in \cite{Reitberger2014b}, we obtain 2D projection maps, spectral energy distributions (SEDs), and integrated flux values for high-energy emission via inverse Compton scattering, bremsstrahlung emission, and neutral-pion-decay. Owing to its anisotropic nature, the inverse Compton component is sensitive to the viewing angle which is determined by the parameters as listed in Section \ref{par}. A schematic view of the topology of the computational domain is shown in Fig. \ref{pion} (left) which depicts the orbital plane with the stars in apastron configuration along with various viewing angles. The red arrow marks the viewing angle which is suggested by observations \citep{Schmutz1997}. Fig. \ref{pion} (right) shows a projected emission map for the neutral pion component if seen along the red line by an observer located on Earth. It becomes clear that the bulk of the emission stems from the apex of the WCR where high-energy proton densities as well as wind plasma densities are highest. We estimate the diameter of the region where $>$99\% of the emission above 100 MeV occurs to be $\sim$700 R$_\odot$. It is smaller for periastron. The analogous plot for bremsstrahlung and inverse Compton components would be empty. Owing to the low maximum energy of the electrons, the leptonic channels do not emit any $\gamma$ rays above 100 MeV. This can be seen in the SEDs in Fig. \ref{SEDs} (middle) which represent the case of apastron. Even if the orbital motion of the system is turned off (as throughout this study), the turbulence of the WCR will lead to minor fluctuations in the SED. This is indicated by the grey-shaded area in Fig. \ref{SEDs} (right) which shows the range of fluctuation for the neutral pion component. We also modeled the system of $\gamma^2$ Velorum using the alternative WR mass-loss rate of $\dot{M}_\mathrm{WR}=8\times10^{-6}$ M$_\odot$ yr$^{-1}$. This 3.75 times lower mass-loss rate leads to a significant reduction of the maximum plasma densities on the WR side of the WCR. Consequently the high-energy proton densities become lower too. A model result that comes close to the measured data points demands an injection ratio of $\eta_p\sim$1 which is highly problematic. However, for $\dot{M}_\mathrm{WR}=3\times10^{-5}$~M$_\odot$~yr$^{-1}$, a far more realistic injection rate of $\eta_p$=10$^{-3}$ is sufficient to reproduce the measured spectrum. The total integrated $\gamma$-ray flux emitted above 100 MeV via the channel of neutral pion decay is $\sim5.6\times10^{-5}$~m$^{-2}$s$^{-1}$, above 10 GeV it is $\sim1.4\times10^{-7}$ m$^{-2}$s$^{-1}$. The bremsstrahlung and inverse Compton channel are only relevant at lower energies. \\

The process of photon-photon absorption in the stellar radiation fields has been taken into account throughout the analysis. This has been done via integrating the optical depths $d\tau$ inside the simulated volume along the line of sight \citep[as detailed in][]{Reitberger2014b}. The emissivities are then lowered by a factor of $e^{-\tau}$. As the difference to a black-body spectrum is small, we assume the stellar photons to be monochromatic. 
 From the simple estimate $E_TE_\gamma\geq(m_ec^2)^2$ one deduces that the process becomes relevant for $\gamma$-ray photons at $E_\gamma\gtrsim$~50~GeV (assuming monochromatic seed photons at $E_T/k_B=56000$ K ). Some of the spectra from neutral pion decay reach maximum energies of $\sim$100 GeV. The effect of photon-photon absorption is clearly visible in the SED of Fig. \ref{SEDs} (centre) as a softening above $\sim~50~$GeV. It has no effect in the energy regime of the \textit{Fermi}-LAT data points.
\\

\subsection{Energetics considerations}
The data presented by \cite{Pshirkov2016} suggests a $\gamma$-ray emissivity of $\sim 5\times10^{31}$ erg s$^{-1}$. This is consistent with our model results which yield a total $\gamma$-ray emissivity of $\sim 5.6\times10^{31}$ erg s$^{-1}$. The presented projected emission maps for neutral pion decay suggest that the $\gamma$ rays are primarily emitted on the WR side of the shock where plasma densities as well as shock velocities are high. The bolometric luminosity of the WR star is $\sim 6.5\times10^{38}$ erg s$^{-1}$. From the given mass-loss rate and terminal velocity one can compute that $\sim 2\times10^{37}$ erg s$^{-1}$ or roughly 3\% of $L_\mathrm{bol}$ are given as the kinetic energy of the wind. However due to radiative braking, $v_\infty$ is not reached in the region relevant for the emission. Extracting density and velocity values from the simulation output, we compute the total kinetic energy available in this region. The sum yields $\sim 5\times10^{35}$ erg s$^{-1}$ or roughly 0.1\% of $L_\mathrm{bol}$. If only one ten-thousandth of this energy budget were used for the emitted spectrum, we could account for the signal we see.

\subsection{Orbital considerations}
The analysis presented in the previous sections is focusing on the orbital state of apastron (stellar separation $d\sim 344 R_\odot$) which is not necessarily representative for the whole orbit. In order to compare matching predictions of our simulations with the measured data, other orbital states have to be taken into account. We therefore selected five orbital phases ($\phi=0,0.1,0.2,0.3,0.4$) for which the same analysis steps as described above for the case of apastron have been carried out. \\
We find a significant decrease of $\gamma$-ray flux towards the periastron state ($\phi$=0, $d=172 R_\odot$). This is due to several factors such as lower shock velocities, a reduced volume of the WCR, and increased Coulomb-losses due to higher densities as the stars draw nearer towards each other. \\
For periastron phase the measured flux levels cannot be reached by tuning the free parameters of injection-rate and normalization of diffusion coefficient (within reasonable ranges). However, a spectrum that lies withing the grey-shaded area shown in Fig. \ref{SEDs} (right) and is thus consistent with the data can easily be found for the orbital state $\phi=0.3$ (with the parameters of $D_0=3\times 10^{-14}$ m$^2 $s$^{-1}$, $\delta$=0.3, and $\eta_p=6\times 10^{-3}$). At this phase, the stellar separation nicely corresponds to the average distance obtained by integrating over the whole orbit. For the same set of parameters, the apastron flux lies considerably above the measured data, periastron flux considerably below. 
\subsection{Nonthermal radio emission}
Having obtained the magnetic field and the electron spectra throughout the simulated volume, we apply the formalisms detailed by \citet{Blumenthal1970} to reach a prediction for the synchrotron emission at radio wavelengths. At 0.2 GHz the resulting spectrum lies $\sim$8 orders of magnitude below the ATCA observations of $\gamma^2$~Velorum. This is due to the rather low chosen surface magnetic field strength of 10$^{-3}$~T and the early cutoff of the electron spectra. For a higher surface magnetic field of 10$^{-2}$~T (while keeping $D_0$, $\delta$, and $\eta_p$ constant at the values stated above for apastron passage), the predicted synchrotron emission is merely one order of magnitude below the ATCA observations. However, the difference between model results and the measured data points increases for higher wavelengths as the simulated synchrotron spectrum decreases much more steeply than the observed signal.\\
An alternative set of parameters in conjunction with smaller inverse Compton losses may bring the predicted nonthermal radio emission in principal agreement with observations. Still, this leaves the composite nature of the radio intensity between thermal and nonthermal radio emission unaddressed, with the former considered being dominant \citep{Chapman1999}. A strong limit is only imposed in a way that our synchrotron prediction cannot supersede the total measured radio intensity, which clearly is not the case.

\section{Summary and Outlook}
\label{disc}

We present the first 3D MHD simulation of the $\gamma^2$ Velorum binary system that successfully reproduces a wide opening angle of the WCR as suggested by observations \citep{Henley2005}. This is achieved by implementing the CAK radiative line acceleration in the limit of strong coupling of stellar and wind parameters. Through severe radiative braking the WR wind is slowed efficiently prior to hitting the WCR resulting in a large opening half-angle of $\sim$72$^\circ$. 

In addition, our model can account for the observed $\gamma$-ray emission from $\gamma^2$ Velorum via diffusive shock acceleration of protons at the WCR and the resulting neutral pion decay emission. Simulating the acceleration of charged particles at the WCR, we obtain proton distributions up to $\sim$ 1 TeV whereas electrons hardly reach 100 MeV. This is mainly due to the significant inverse Compton and synchrotron losses at the WCR. Maximum proton energies and flux levels are determined by our choice of diffusion coefficient, diffusion index and proton injection ratio. To obtain approximate agreement with measured $\gamma$-ray spectra, we choose $D_0=8\times 10^{-14}$ m$^2 $s$^{-1}$, $\delta=0.3$, and $\eta_p=10^{-3}$ for the apastron phase of the system. Most orbital phases can be modeled to fit the data by changing these parameters within a reasonable range. For phases around periastron this is no longer possible. This suggests the presence of variability on orbital timescales the non-observation of which can be accounted for by the low statistics.

2D projection maps of the $\gamma$-ray emission region show that the bulk of the emission due to neutral pion decay is clearly confined to a region around the apex of the WCR of a diameter of roughly $\sim700$ R$_\odot$. A larger computational domain would add little to the total $\gamma$-ray emission.

The observed data by \cite{Pshirkov2016} can only be attained by $\gamma$ rays of hadronic origin as the leptonic components are far too weak due to the low energy of the electrons. Of the two mass-loss rates for the WR star found in literature only the higher one is fit to reproduce the observed spectra with a realistic choice for the proton injection ratio. $\dot{M}_\mathrm{WR}=8\times10^{-6}$ M$_\odot$yr$^{-1}$ would require $\eta_p=1$ which is considered unrealistically large. Owing to the turbulent WCR the output spectra do not converge to a definite flux level but rather fluctuate above and below the observed data. \\

This work lays the base line for further investigation of the $\gamma^2$ Velorum via numerical simulation. The results on particle acceleration and $\gamma$-ray emission focus on the orbital phase of apastron but also show that similar agreement with observations can be achieved for other orbital states. 
In the near future we plan to compare numerical simulations of $\gamma$-ray emission in $\gamma^2$~Velorum with the much brighter $\gamma$-ray source of $\eta$ Carinae and the yet undetected but suspected $\gamma$-ray source WR~140. If we apply the same parameters regarding diffusion and particle injection as for $\gamma^2$~Velorum -- can simulations account for the fact that one source is bright while the other remains dark on the \textit{Fermi} $\gamma$-ray sky? We hope to provide an answer to this question in the near future.

\acknowledgments
The computational results presented have been achieved (in part) using the HPC infrastructure of the University of Innsbruck. A.R. acknowledges financial support from the Austrian Science Fund (FWF), project P~24926-N27.

\bibliographystyle{aasjournal}

\appendix

\section{On 1$^\mathrm{st}$-order Fermi-acceleration}
In our modelling efforts we use the cosmic-ray transport equation following \citet{Parker1965}:
\begin{equation}
	\label{EqTransportPhaseSpace}
	\frac{\partial f}{\partial t}
	-
	\nabla \cdot \left(
  D \nabla f\right)
  +
  \vec{u}
  \cdot
  \nabla f
  -
  \frac{1}{3} \left(\nabla \cdot \vec{u}\right)
  \frac{\partial f}{\partial \ln p}
  =
  q_f(\vec{x},\vec{p}),
\end{equation}
where $f$ phase space density of the energetic particles depending on time $t$, position $\vec{x}$, and momentum $\vec{p}$ where $p=|\vec{p}|$. Additionally, $D$ is the magnitude of spatial diffusion, $\vec{u}$ is the advection velocity, and $q_f$ the source term. The measurable quantity is the particle flux $j = 4\pi p^2 f$. By using:
\begin{align}
	4\pi p^2 \left(\nabla \cdot \vec{u}\right) \frac{\partial f}{\partial \ln p}
	&=
	4 \pi \left(\nabla \cdot \vec{u}\right) \frac{\partial}{\partial p}\left(p^3 f\right)
	-
	4\pi \left(\nabla \cdot \vec{u}\right) 3 p^2 f
	\nonumber\\
	&=
	\left(\nabla \cdot \vec{u}\right)\frac{\partial}{\partial p}\left(p j\right) - 3\left(\nabla \cdot \vec{u}\right) j
\end{align}
the transport equation for the flux $j$ is found to be:
\begin{equation}
	\label{EqTransportFlux}
	\frac{\partial j}{\partial t}
	-
	\nabla \cdot \left(
  D \nabla j\right)
  +
  \nabla \cdot \left(
  \vec{u}
  j
  \right)
  -
  \frac{1}{3} \left(\nabla \cdot \vec{u}\right)
  \frac{\partial}{\partial  p}\left(p j\right)
  =
  q(\vec{x},p),
\end{equation}
where the source term $q$ relates to the one in Eq. \ref{EqTransportPhaseSpace} by $q= 4\pi p^2 q_f$.
Note that this is analogous to Eq. \ref{D_eq} with the minor differences that the latter is expressed in terms of energy $E$, rather than absolute momentum $p$, includes energy losses $\dot{E}_\mathrm{loss}$, assumes isotropic diffusion, and specifies the source term to an injection at a given energy $E_0$.  

\subsection{Diffusive shock acceleration}
The transport equations \eqref{EqTransportPhaseSpace}, \eqref{EqTransportFlux} have been investigated by several authors \citep[see][]{Krymskii1977,Axford1977,Blandford1978} in the context of particle acceleration at an infinitely extended shock perpendicular to the magnetic field in the plasma. By assuming constant upstream and downstream velocity, matching the respective solutions of the transport equation at the position of the shock shows that particles injected at the shock obtain a power-law spectrum.

For the same scenario it is also possible to investigate the temporal change of the energy-dependent particle flux. This is, e.g., extensively discussed in  \citet{Drury1983}, where the acceleration rate of the energetic particles can be found by applying a Laplace transform to the transport equation. Here, we will recapture the discussion from that study for the specific case of our numerical setup.



\subsection{Acceleration Rate}
To investigate the acceleration rate, we consider an infinitely extended shock with homogeneous upstream and downstream plasmas. Additionally assuming spatially constant diffusion, \eqref{EqTransportFlux} can be used in the form \citep[see][]{Drury1983}:
\begin{equation}
	\label{EqTransport1D}
	\frac{\partial j}{\partial t}
	-
	D \frac{\partial^2 j}{\partial x}
	+
	\frac{\partial}{\partial x} \left(
	u
	j
	\right)
	-
	\frac{1}{3} \frac{\partial u}{\partial x}
	\frac{\partial}{\partial  p}\left(p j\right)
	=
	q(x,p),
\end{equation}
Assuming injection to be localised at the shock at a given momentum ($q(x,p) = q_0 \delta(p-p_0) \delta(x-x_{\text{shock}})$ this equation becomes considerably simpler in the homogeneous upstream or downstream medium:
\begin{equation}
	\label{EqTransp1DUpDown}
	\frac{\partial j}{\partial t}
	-
	D \frac{\partial^2 j}{\partial x}
	+
	u_i \frac{\partial j}{\partial x}
	=
	0,
\end{equation}
where we explicitly used the constant velocity in the upstream $u_i = u_u$ and the downstream $u_i = u_d$ regions.

The Laplace transform:
\begin{equation}
	J(x,p,s) = \int \limits_0^{\infty} e^{-st} j(t,x,p) dt
\end{equation}
applied to Eq. \eqref{EqTransp1DUpDown} gives:
\begin{equation}
	\label{EqTranspLaplace}
	s J + u_i \frac{\partial J}{\partial x} - D \frac{\partial^2 J}{\partial x^2}
	=
	0,
\end{equation}
where $j(t=0,x,p) = 0$. Since the particles originate and are accelerated at the shock, $J=0$ for $x\to \pm \infty$. Thus, the solution of Eq. \eqref{EqTranspLaplace} is:
\begin{equation}
	\label{EqSpatialDist}
	J \propto e^{\beta_i x}
	\qquad\text{with}\qquad
	\beta_i
	=
	\frac{u_i}{2 D}\left(1 \pm \left(1 + \frac{4 D s}{u_i^2}\right)^{1/2}\right)
\end{equation}
with $(+)$ for the upstream and $(-)$ for the downstream case.

Now, the downstream and upstream solutions need to be matched at the shock position. First, the flux needs to be continuous there \citep[see][]{Drury1983}, implying:
\begin{equation}
	J_u(x \to x_{\text{shock}})
	=
	J_d(x \to x_{\text{shock}})
\end{equation}

 Secondly, an additional constraint is found by integrating Eq. \eqref{EqTransport1D} from a position just upstream to one just downstream of the shock.
Considering that $j$ is also continuous at the shock, this leads to:
\begin{align}
	-
	D \left.\frac{\partial j}{\partial x}\right|_d
	+
	D \left.\frac{\partial j}{\partial x}\right|_u
	+
	u_d j
	-
	u_u j
	-
	\frac{1}{3}\frac{\partial (p j)}{\partial p}
	\left(u_d - u_u\right)
	&=
	\nonumber\\
	-
	D \left.\frac{\partial j}{\partial x}\right|_d
	+
	D \left.\frac{\partial j}{\partial x}\right|_u
	+
	\frac{2}{3}
	\left(
	u_d j
	-
	u_u j
	\right)
	-
	\frac{1}{3}p\frac{\partial j}{\partial p}
	\left(u_d - u_u\right)
	&=
	p q_0 \delta (p-p_0)
\end{align}
Taking the Laplace transform of this at the position of the shock leads to:
\begin{align}
	\label{EqLaplaceAtShock}
	D \left(\beta_u - \beta_d\right) J_0
	-
	\frac{2}{3}(u_u - u_d) J_0
	+
	\frac{1}{3}p\frac{\partial J_0}{\partial p}
	\left(u_u - u_d\right)
	&= 
	\int\limits_0^{\infty}
	q_0 \delta (p-p_0) e^{-st} dt
	\nonumber\\
	&= 
	\frac{p q_0 \delta (p-p_0)}{s}
\end{align}
where $J_0 = J(s,0,p)$ is the Laplace transform at the shock. In analogy to the derivation in \citet{Drury1983} we introduce:
\begin{equation}
	A_i(s) = \left(\left(1 + \frac{4 D s}{u_i^2}\right)^{1/2} - 1\right)
\end{equation}
With this, it can be shown that Eq. \eqref{EqLaplaceAtShock} can be written as:
\begin{equation}
	\frac{1}{2}
	\left(
	u_u A_u(s) + u_d A_d	(s)
	\right)
	J_0
	+
	\frac{1}{3}
	\left(u_u + 2 u_d\right) J_0
	+
	\frac{1}{3}p\frac{\partial J_0}{\partial p}
	\left(u_u - u_d\right)
	= 
	\frac{p q_0 \delta (p-p_0)}{s}
\end{equation}
Considering that the inhomogeneity is only non zero at $p=p_0$ yields the solution:
\begin{equation}
	J_0(s,p_1)
	=
	\frac{3 q_0}{s(u_u - u_d)}
	\left(\frac{p_1}{p_0}\right)^{-\frac{u_u + 2 u_d}{u_u - u_d}}
	\exp
	\left(
	-\int \limits_{p_0}^{p_1}
	\frac{3}{2}
	\frac{u_u A_u(s) + u_d A_d(s)}{u_u - u_d} \frac{dp}{p}
	\right)
\end{equation}
From this the time-dependent particle flux at the shock follows from the related Bromwich integral:
\begin{equation}
	j_0(t,p) = j(t,x=0,p)
	=
	\frac{1}{2\pi\imath}
	\int\limits_{\eta - \imath \infty}^{\eta + \imath \infty}
	J_0(s,p) e^{ts} ds,
\end{equation}
where $\eta \to 0$ since the largest value of $s$ for any singularity is $s=0$, here. The asymptotic behaviour is found from the contribution of the residual with the largest real part, which is the simple pole at $s=0$ leading to:
\begin{equation}
	j_0(t\to\infty,p)
	=
	j_0(\infty,p)
	=
	\frac{3 q_0}{u_u - u_d}
	\left(\frac{p_1}{p_0}\right)^{-\frac{u_u + 2 u_d}{u_u - u_d}}
	\propto
	p_1^{-a}
	\qquad\text{with}\qquad
	a = \frac{c_r+2}{c_r-1}	
\end{equation}
This leads to the usual result that for a compression ratio of $c_r\to4$ the spectral index becomes $a\to2$. Additional consideration of the spatial dependence upstream of the shock in the limit $t\to\infty$ shows that Eq. \eqref{EqSpatialDist} leads to
\begin{equation}
	\label{EqExpDecayInfty}
	j \propto e^{\beta_u x}
	\qquad
	\text
	\qquad
	\beta_u = \frac{u_u}{D},
\end{equation}
representing the usual exponential decrease in the upstream direction.

In the general case the inverse transform reads:
\begin{align}
	j_0(t,p) 
	&=
	\frac{1}{2\pi\imath}
	\int\limits_{\eta - \imath \infty}^{\eta + \imath \infty}
	\frac{3 q_0}{s(u_u - u_d)}
	\left(\frac{p_1}{p_0}\right)^{-\frac{u_u + 2 u_d}{u_u - u_d}}
	\exp
	\left(
	-\int \limits_{p_0}^{p_1}
	\frac{3}{2}
	\frac{u_u A_u(s) + u_d A_d(s)}{u_u - u_d} \frac{dp}{p}
	\right) e^{ts} ds
	\nonumber\\
	&=
	\frac{1}{2\pi\imath}
	\int \limits_0^t
	dt'
	\int\limits_{\eta - \imath \infty}^{\eta + \imath \infty}
	\frac{3 q_0}{u_u - u_d}
	\left(\frac{p_1}{p_0}\right)^{-\frac{u_u + 2 u_d}{u_u - u_d}}
	\exp
	\left(
	-\int \limits_{p_0}^{p_1}
	\frac{3}{2}
	\frac{u_u A_u(s) + u_d A_d(s)}{u_u - u_d} \frac{dp}{p}
	\right) e^{t's} ds
	\nonumber\\
	&=
	\frac{1}{2\pi\imath}
	\int \limits_0^t
	dt'
	\int\limits_{\eta - \imath \infty}^{\eta + \imath \infty}
	j_0(\infty,p)
	\exp
	\left(
	t' s
	-\int \limits_{p_0}^{p_1}
	\frac{3}{2}
	\frac{u_u A_u(s) + u_d A_d(s)}{u_u - u_d} \frac{dp}{p}
	\right) ds
\end{align}
Apparently, this can be expressed as:
\begin{equation}
	j_0(t,p)
	=
	j_0(\infty,p)
	\int \limits_0^t
	\phi(t') dt'
\end{equation}
where
\begin{equation}
	\phi(t)
	=
	\frac{1}{2\pi\imath}
	\int\limits_{\eta - \imath \infty}^{\eta + \imath \infty}
	e^{ts - h(s)} ds
\end{equation}
and
\begin{equation}
	\label{EqDefh}
	h(s) = \int \limits_{p_0}^{p_1}
	\frac{3}{2}
	\frac{u_u A_u(s) + u_d A_d(s)}{u_u - u_d} \frac{dp}{p}
\end{equation}
From the definition of $\phi(t)$ it follows that:
\begin{equation}
	\label{EqIntPhi}
	\int\limits_0^{\infty}
	\phi(t) e^{-t s} dt
	=
	e^{-h(s)}
\end{equation}
Thus, $\phi(t)$ can be interpreted as the acceleration time distribution for given $p_0$ and $p_1$ for $h=0$, where the distribution is normalised \citep[see][]{Drury1983}. From the definition of $h$ in Eq. \eqref{EqDefh} this represents the case $s=0$. Thus, differentiating Eq. \ref{EqIntPhi} with respect to $s$ and setting $s=0$ gives:
\begin{equation}
	-\int \limits_0^{\infty} t dt
	=
	- \left . \frac{\partial}{\partial s}h(s)\right|_{s=0}
\end{equation}
This leads to the following expression for the mean acceleration time:
\begin{equation}
	\left<t\right>
	=
	\left.
	\frac{\partial}{\partial s}h(s)\right|_{s=0}
	= 
	\frac{3}{u_u - u_d}
	\int \limits_{p_0}^{p_1}
	D
	\left(
	\frac{1}{u_u} + \frac{1}{u_d}
	\right)
	\frac{dp}{p}
\end{equation}
This finally shows that the rate of momentum gain is given as:
\begin{equation}
	\label{EqAccTime}
	\dot p
	=
	\frac{p}{t_{acc}}
	\qquad
	\text{with}
	\qquad
	t_{acc} = \frac{3}{u_u - u_d}
	D
	\left(
	\frac{1}{u_u} + \frac{1}{u_d}
	\right)
	=
	\frac{3 D}{u_u^2}\frac{(r+1)r}{r-1}
\end{equation}
being the usual expression used in many studies of diffusive shock acceleration.

\subsection{A direct estimate of the acceleration rate}
The acceleration rate can also be found by investigating the term reflecting the adiabatic energy changes in the transport equation. Considering only the first and the fourth term of Eq. \eqref{EqTransportFlux} this part of the transport equation can be viewed as an advection equation in momentum with an advection velocity:
\begin{equation}
	v_p = \frac{1}{3} \left(\left(\nabla \cdot \vec{u}\right) p\right)
\end{equation}
representing the rate of momentum gain. When evaluating this expression at the position of the shock, the velocity divergence is:
\begin{equation}
	\nabla \cdot \vec{u} = \frac{u_d - u_u}{\Delta x}
	=
	\frac{1}{\Delta x} \frac{u_d - u_u}{u_d} u_d
	=
	\frac{1}{\Delta x} (1-r) u_d \frac{u_u}{u_u}
	=
	\frac{1}{\Delta x} \frac{1-r}{r} u_u
\end{equation}
The relevant length-scale can, additionally, be found from physical arguments. First, we noticed the exponential decrease of the flux in the upstream direction. According to \citet{Blasi2004} the point where the flux has decreased by a factor $e$ can be interpreted as the position, where the particles on average change direction. According to Eq. \eqref{EqExpDecayInfty} this happens at a distance:
\begin{equation}
	x_u = (\beta_u)^{-1} = \frac{D}{u_u}
\end{equation}
upstream of the shock. On the downstream side such an estimate is not trivial anymore, since the distribution becomes spatially constant in the limit $t\to\infty$. \citet{Drury1983}, however, shows that the mean residence time upstream and downstream of the shock follow the same analytical expression, thus motivating:
\begin{equation}
	x_d = \frac{D}{u_d}
\end{equation}
Using $\Delta x = x_u + x_d$ then yields:
\begin{equation}
	\nabla \cdot \vec{u}
	=
	\frac{1}{D}
	\frac{1-r}{r} u_u
	\frac{1}{\frac{1}{u_u} + \frac{1}{u_d}}
	=
	\frac{1}{D}
	\frac{1-r}{r} u_u	
	\frac{u_u u_d}{u_u + u_d}
	=
	\frac{1}{D}
	\frac{1-r}{r} u_u^2	
	\frac{1}{r + 1}
\end{equation}
Thus, we find for the momentum rate of change:
\begin{equation}
	v_p = \frac{\partial p}{\partial t}
	=
	\frac{u_u^2}{3 D}
	\frac{1-r}{r(r + 1)}p
\end{equation}
with a resulting time scale that is identical to the one found in Eq. \eqref{EqAccTime}. 

Additionally, this shows the limits of the numerical representation of the transport equation. A shock computed in a numerical simulation has a finite thickness on the order of a few numerical cells. As long as this thickness is below $\Delta x$ as computed above the accelerated particles can be expected to experience the shock as a discontinuity, leading to the same acceleration rate as for an actual discontinuity. Deviations are only to be expected when the particle's diffusion is so low that  $\Delta x$ becomes smaller than the simulated shock thickness.

\end{document}